\documentclass{aa} 
\usepackage{graphicx}
\usepackage{siunitx}
\usepackage{breqn}
\usepackage{float}
\usepackage{txfonts}
\DeclareMathOperator{\e}{e}
\graphicspath{{fig/}}
\begin{document}

   \title{Exoplanets detection limits using spectral cross-correlation with spectro-imaging}

   \subtitle{An analytical model applied to the case of ELT-HARMONI}

   \author{A. Bidot,
          \inst{1}
          D. Mouillet
          \inst{1}
          \and
          A. Carlotti,
          \inst{1}
          }

   \institute{Institute of Planetology and Astrophysics (IPAG), University of Grenoble Alpes,
              414 rue de la Piscine, Saint-Martin d'Hères\\
              \email{alexis.bidot@univ-grenoble-alpes.fr}}

   \date{Received February, 2023; accepted }

  \abstract
{The combination of high-contrast imaging and medium to high spectral resolution spectroscopy offers new possibilities for the detection and characterization of exoplanets. 
The so-called molecular mapping technique makes use of the difference between the planetary and stellar spectra.
Whereas traditional high-contrast imaging post-processing techniques are quickly limited by speckle noise at short angular separation, molecular mapping efficiently suppresses speckles offering new detection possibilities.
}
{Molecular mapping performance depends on multiple parameters such as the star magnitude, the adaptive optics residual halo, the companion spectrum, the telluric absorption, as well as the telescope and instrument properties. 
Exploring such a parameter space through end-to-end simulations to predict potential science cases and to optimize future instruments designs is then very time-consuming and makes it difficult to draw simple conclusions. 
We propose to quantify the signal of interest and the noise that propagates in molecular mapping, with explicit dependencies upon the stellar, planetary, and instrument main parameters, in order to define an efficient methodology for such an analysis.}
{ Explicit expressions of the estimates of molecular mapping signal and noise are derived, and they are validated through comparisons with end-to-end simulations. They provide an understanding of the instrumental dependencies, and help to discuss optimal instrumental choices with regard to the targets of interest. They are applied in the case of the ELT/HARMONI integral field spectrograph, as an online tool predicting the contrast performance in various observational cases.}
{We confirm the potential of molecular mapping for high-contrast detections, especially for cool planets at short separations. We provide guidelines based on quantified estimates for design trade-offs of future instruments. We discuss the planetdetection performances of the HARMONI observing modes, with spectral resolutions from 3000 to 17000, and different corresponding spectral bandwidths in the near infrared. While they nicely cover the appropriate requirements for high detection capability of warm exoplanets, the high-contrast mode of HARMONI would benefit from a transmission extended down to J band. A contrast of a few $10^{-7}$ at 50 mas should be within reach on bright targets in photon noise regime with molecular mapping.
}
{}

   \keywords{ Instrumentation: high angular resolution -- Methods: data analysis --  Techniques: imaging spectroscopy -- Planets and satellites: detection}

   \maketitle
%
%-------------------------------------------------------------------
%%%%%%%%%%%%%%%%%%%%%%%%
\section{Introduction}
%%%%%%%%%%%%%%%%%%%%%%%%

The spectral information that is carried by the light reflected of, or emitted by an exoplanet is of primary importance for its study. It makes it possible to probe the physical properties and the composition of the atmosphere ; this is a proxy to infer the formation and evolution processes \citep[][]{Oberg_2016, Molliere_2022}, and a key information to assess their habitability \citep[][]{turbet_2018}, and the potential existence of life \citep[][]{J_Wang_2018}.
The brightness of the host star, and the huge star-to-planet flux ratio induce noise and systematics on the planet's signature, which are two major challenges to these observations.

Such detection and characterization of the exoplanetary light can be attempted with or without angularly resolving the planet itself, and, in each case, considering the total, broadband planetary flux, or trying to focus on some specific spectral features.  
Each case involves different dominant limitations that lead to different potential ultimate performances, as we briefly review hereafter. 

Starting from the spatially unresolved case, in the favorable configuration of transiting planets, the planetary signal can be extracted thanks to temporal differential spectro-photometry \citep[][]{Tinetti2018}, i.e., by observing the planet before, during, and after the transit. The performance of this technique is ultimately limited by systematics related to measurement stability, and by the limited time window associated with transits (once every orbit). 

The spectral diversity between the star and the planet signal, and/or the use of specific spectral signatures, can be used to get rid of these instrumental systematics, to the point that even non-transiting planets can be detected and characterized without spatially resolving them. This enabled the detection and characterization of $\beta$-Pic b observed with VLT/CRIRES \citep[][]{cross_corr_snellen_2014}, for instance.
The planetary signal detection remains strongly limited by the photon noise associated with the huge stellar flux.

This noise is dramatically reduced if the exoplanet and the star can be spatially separated from each other. It can be further reduced if the star's diffracted signal at the location of the planet is attenuated using coronagraphic techniques. As for unresolved characterization, differential signal extraction can be performed without using any prior knowledge of the exoplanet spectrum.
The key limitation is associated with systematics on images, in particular speckle noise, in spite of various studies and huge efforts that have been made to handle it (such as improved AO correction methods, novel focal plane sensing techniques, improved signal extraction algorithms that use data diversity). One way to get rid of the confusion between the speckles in the stellar halo and the planetary signal is the interferometric coherence difference, as used in particular in the ExoGRAVITY program \citep[][]{Lacour2020}; this approach is telescope-time expensive, however, since it requires the 4 UTs of the VLT. In addition, it has a low global transmission. As said before, another way to deal with these systematics is to use the spectral information to look for specific spectral features, over the field of view. This technique referred to as molecular mapping. 
This requires a high enough spectral resolution (which is discussed below) to distinguish the absorption features of interest from the speckle-induced spectrum modulation. The integral field spectrographs (IFS) used with high-contrast imaging instruments such as VLT/SPHERE \citep[][]{Beuzit2019}, GPI \citep[][]{GPI_instru}, Subaru/SCExAO \citep[][]{ScexAO} focus on spatial information, and have a limited spectral resolution (R=50). However, molecular mapping has begun to be demonstrated with higher resolution IFS and spectrographs that were not originally dedicated to high-contrast: e.g. with VLT/SINFONI (R=5000) \citep[][]{Hoeijmakers2018}, Keck/OSIRIS (R=4000) \citep[][]{ruffio_2019}, and VLT/MUSE (R=3450) \citep[][]{Haffert2019}. High-Dispersion Coronagraphy (HDC) has also been proposed and designed to couple high-resolution spectrographs with high-contrast instruments, and observe single, diffraction-limited sources, allowing to characterize exoplanets whose positions are already known: KPIC (R=37000) \citep[][]{Delorme2018,Wang2021,XUAN_2022}, HiRISE (R=100000) \citep[][]{Vigan_2022}, VIPA (R=80000) \citep[][]{VIPA_2022}.\\

There is a full parameter space to explore when following this approach, with various balances and trade-offs between the spectral information content, noise regimes, observation context (ground vs. space), as well as star and exoplanet types. 
We should emphasize the fact that, in practice, when designing an instrument, the main design choices usually favor some capabilities at the expense of others. A classical example is, for instance, the necessary trade-off, for an IFS with a given detector size, between its field-of-view (FoV), spectral bandwidth, and spectral resolution. There is certainly not a single, universal sweet spot, and the choice depends on various parameters that include the telescope specifications, the image quality, the stellar type and brightness, the total observing time, the exoplanet contrast, projected separation, spectral properties, etc., and the chosen priority given to a given spectral signature. 
Some IFS using this high spectral resolution approach are already operational but do not necessarily have the optimal observing modes for the detection of exoplanets (VLT/ERIS \citep[][]{ERIS}, VLT/MUSE, Keck/OSIRIS, JWST/MIRI \citep[][]{JWST_molmap, malin}). Others are currently being designed with these trade-off considerations in mind (SPHERE+ MedRES \citep[][]{SPHERE+}, ELT/HARMONI, ELT/METIS \citep[][]{METIS_2020}), or are planned in the more distant future (ELT/PCS) \citep[][]{PCS}.
Such discussions on the instrument trade-offs and on the different high-resolution instrument modes have already been addressed in some specific cases, such as  space-based very-high-contrast instruments \citep[][]{WANG_JI_2017}, or for coupling VLT/SPHERE with CRIRES+ \citep[][]{Otten_2021}. This will need to be extended, and to cover a variety of future instruments. \\

We argue that a new performance analysis approach is needed to explore this parameter space, and we propose for that to use a semi-analytical tool that we developed. Until now, a huge effort has been made to develop end-to-end simulation tools to estimate the performance of various high-contrast imagers; these simulations are necessary to model and to estimate the impact of complex effects - such as speckle noise, instrumental instabilities, or correlations generated by the detector - on the final achievable contrast, with various post-processing techniques. A strong drawback of this computationally heavy approach is the difficulty to explore wide parameter spaces, and to fully trace and understand the role of each of the various parameters that contribute to the performance budget, however. Our approach presented here proposes an answer to this difficulty.\\

We first recall in Section \ref{molmapsection} the principles of molecular mapping, and how different types of noise limit its efficiency. We verify our analytical model with comparisons with end-to-end simulations in Section \ref{section:validation}. Section \ref{trade_off} details the bandwidth vs. spectral resolution trade-off, and how tellurics, and rotational broadening may influence it. We then consider in Section \ref{section:application_HARMONI} the specific case of the ELT/HARMONI instrument to first compare the performance derived through an end-to-end simulation tool and the one derived with our semi-analytical tool, and then use the latter to discuss the interest of the different observing modes of this instrument with respect to different types of planets. Finally, a conclusion is drawn in Section \ref{conclusion}.

%%%%%%%%%%%%%%%%%%%%%%%%%%%%%%%%%%%%%%%%%%%%%%%%%%%%
\section{Molecular Mapping information and noise} \label{molmapsection}
%%%%%%%%%%%%%%%%%%%%%%%%%%%%%%%%%%%%%%%%%%%%%%%%%%%%

\subsection{Molecular mapping core principle}

We consider the observational information of spectro-imagers, after the first steps of data reduction and calibration, in the form of a 3D collected flux $S$, expressed as the measured integrated flux as a function of the wavelength $\lambda$ and spaxel position ($x$,$y$). It can be decomposed into several components when the observed scene is a stellar halo as in Eq.\ref{composants}.
\begin{eqnarray}
    S(\lambda,x,y) &=& \gamma_{\mathrm{atm}}(\lambda)\cdot\left(M_{\mathrm{speckle}} (\lambda,x,y).S_{\mathrm{star}} (\lambda) + S_{\mathrm{planet}} (\lambda,x,y)\right) \nonumber \\ \label{composants}
    &\, & + n(\lambda,x,y) \\  \label{eq:noise_breakdown}
n(\lambda,x,y) &=& n_{\mathrm{halo}}(\lambda,x,y) + n_{\mathrm{bkgd}}(\lambda) + n_{\mathrm{RON}}   
\end{eqnarray}
with $\gamma_{\mathrm{atm}}$ the atmosphere transmission, $S_{\mathrm{star}}$ the star spectrum, $S_{\mathrm{planet}}$ the planet PSF as a function of wavelength, and $n$ the noise. The sky background spectrum does not appear here, as we assume that it has been perfectly subtracted from the data. The total noise $n$ is further broken down in Eq.\ref{eq:noise_breakdown} into several standard noise sources: the stellar photon noise $n_{\mathrm{halo}}$, the background noise $n_{\mathrm{bkgd}}$, which is the photon noise coming from the sky emission, the telescope, and the instrument, and the readout noise of the detector $n_{\mathrm{RON}}$.
$M_{\mathrm{speckle}}$ describes the stellar PSF over the FoV and depends on the wavelength. It is normalized for each spectral channel such that $\iint M(\lambda,x,y) dx dy = 1$. Spectra quantities ($S_{\mathrm{star}}$, $S_{\mathrm{planet}}$, and consequently also $S$) are expressed in detected photo-electrons, integrated over the total integration time for each spectral channel. Therefore, those quantities consider the intrinsic star and planet apparent magnitudes, and also the telescope and instrument transmissions. %The term $S_{\mathrm{planet}}$ includes the dependencies over ($x$,$y$) which represent the apparent spatial extent of the companion (i.e., the system point-spread function (PSF) at each wavelength). 

The PSF and the speckle pattern scale with the wavelength as a first order approximation. Thus, the scaling of the speckle pattern generates a low frequency modulation of the stellar spectrum, in an uncontrolled manner, depending on the location in the field of view. In order to isolate the planetary spectrum from the stellar spectrum, the stellar continuum modulated by the speckles must be estimated and subtracted. This step can be interpreted as high-pass filtering to isolate the uncontaminated frequencies of the planetary spectrum. These frequencies correspond to the absorption lines of the planetary spectrum. To do so, we used the same preprocessing routine as in TexTris (\citet{Petrus2021}). It can be summarized in a few steps:

Firstly, models of the observed star spectrum – $\hat{S}_{\mathrm{star}}$- and tellurics absorption – $\hat{\gamma}_{\mathrm{atm}}$ - are chosen. These can either be models derived from libraries such as PHOENIX for the star, and SkyCalc for tellurics, or directly estimated from the data using the brightest spaxels in the FoV uncontaminated by the companion(s) (i.e. the spaxels corresponding to the core of the stellar PSF: $\hat{\gamma}_{\mathrm{atm}}(\lambda)\cdot \hat{S}_{\mathrm{star}}(\lambda)$ = $<S(\lambda)>_{xy}$).

These models are needed to estimate the speckle modulations in the FoV, so that stellar light can be subtracted. Indeed, for each spaxel, we estimate the modulation by low-pass filtering the ratio between the spaxels and a model of the spectrum of the star, multiplied by the transmission of our atmosphere:

\begin{dmath}
     \hat{M}_{\mathrm{speckle}}(\lambda,x,y)  = \Bigg[\frac{S(\lambda,x,y)}{\hat{\gamma}_{\mathrm{atm}}(\lambda).\hat{S}_{\mathrm{star}}(\lambda)}\Bigg]_{\mathrm{LF}}
     \label{specmod_1}
\end{dmath}

We use the notation $[.]_{\mathrm{LF}}$ to represent the low-pass filtering operation. Here it is just a convolution between a spectrum and a Gaussian kernel with a width $\sigma_c$, homogeneous to nanometer, and defined by the user. This width $\sigma_c$ will determine the frequency cut-off - $f_c$ - of the filter (homogeneous to $nm^{-1}$), and it can be also understood as a spectral resolution cut-off - $R_c$ - as in Eq. \ref{eq:freq_coupure} (see Appendix \ref{appendixA}):

\begin{equation}
\begin{aligned}
    \label{eq:freq_coupure}
    f_c &= \frac{\sqrt{\ln{2}}}{\sqrt{2}\pi\sigma_c}, &\quad R_c &= \lambda_0 \cdot f_c 
\end{aligned}
\end{equation}

where $\lambda_0$ is the central wavelength of the observation bandwidth.
The optimal resolution cut $R_c$  of this filter is discussed later in subsection \ref{speckle_noise}. The high-frequency content of a spectrum $[S]_{\mathrm{HF}}$ will represent the difference $S - [S]_{\mathrm{LF}}$.

Assuming that the model of the star spectrum with the tellurics absorption is perfect and injecting Eq.\ref{composants} into Eq.\ref{specmod_1} returns the estimated modulation function:

\begin{dmath}
     \hat{M}_{\mathrm{speckle}}(\lambda,x,y)  
     = M_{\mathrm{speckle}}(\lambda,x,y) + \Bigg[\frac{S_{\mathrm{planet}}(\lambda,x,y)}{\hat{S}_{\mathrm{star}}(\lambda)}\Bigg]_{\mathrm{LF}} +
     \Bigg[\frac{n(\lambda,x,y)}{\hat{\gamma}_{\mathrm{atm}}(\lambda).\hat{S}_{\mathrm{star}}(\lambda)}\Bigg]_{\mathrm{LF}}
     \label{specmod_2}
\end{dmath}

Note that Eq. \ref{specmod_2} should be an approximation in real life, since the model of the spectrum of the star and telluric absorptions cannot be perfect.

The modulated stellar spectrum must be removed from the spaxels. We note that the planet's continuum will be removed, as the estimated modulation $\hat{M}_{\mathrm{speckle}}(\lambda,x,y)$ will take it into account. $S_{\mathrm{res}}(\lambda,x,y)$, the residual - spectrally high-pass filtered - signal after stellar component subtraction is expressed as:
\begin{dmath}
    S_{\mathrm{res}}(\lambda,x,y) = S(\lambda,x,y) - \hat{M}_{\mathrm{speckle}}(\lambda,x,y)\cdot\hat{\gamma}_{\mathrm{atm}}(\lambda)\cdot \hat{S}_{\mathrm{star}}(\lambda)
    = \gamma_{\mathrm{atm}}(\lambda)\cdot\Bigg([S_{\mathrm{planet}}]_{\mathrm{HF}}(\lambda,x,y) - [\hat{S}_{\mathrm{star}}]_{\mathrm{HF}}(\lambda)\cdot\Bigg[\frac{S_{\mathrm{planet}}(\lambda,x,y)}{\hat{S}_{\mathrm{star}}(\lambda)}\Bigg]_{\mathrm{LF}}\Bigg) + n'(\lambda,x,y)
    \label{residuals}
\end{dmath}
where $\displaystyle n'(\lambda,x,y) = n(\lambda,x,y) - [n(\lambda,x,y)]_{\mathrm{LF}} - \frac{[\gamma_{\mathrm{atm}}\cdot \hat{S}_{star}(\lambda)]_{\mathrm{HF}}}{[\gamma_{\mathrm{atm}}\cdot \hat{S}_{star}(\lambda)]_{\mathrm{LF}}} \cdot [n(\lambda,x,y)]_{\mathrm{LF}}$. 
\newline

In practice, $n'(\lambda,x,y)$ is approximately equal to $n(\lambda,x,y)$, and we will use this approximation in the rest of the paper. The detailed calculation and discussion can be found in Appendix \ref{appendixB}.

\begin{figure*}
    \centering
    \includegraphics[width=18cm]{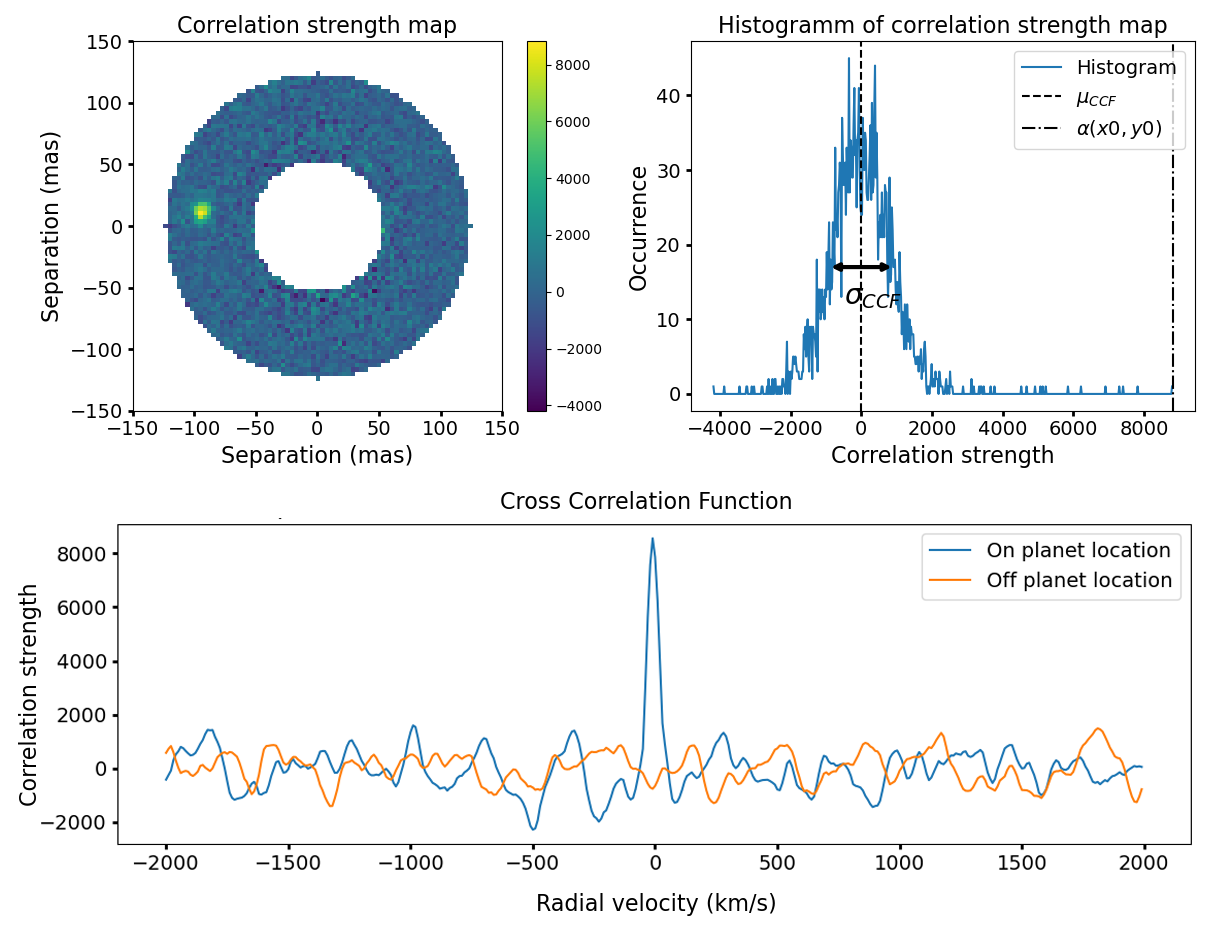}
    \caption{\label{SNRgauss}Top left panel: 2D correlation map processed on simulated PSF with a fake planet injected. The colorbar represents the correlation strength between spaxels and the template. Top right panel: histogram of the correlation strength map, showing the distribution of the correlation values in the annulus. $\mu_{\mathrm{CCF}}$ and $\sigma_{\mathrm{CCF}}$ denotes respectively the mean and the standard deviation of these values. $\alpha(x0, y0)$ is the correlation at the location of the planet. Bottom panel: cross-correlation function on (blue) and off (orange) the planet location.}
\end{figure*}

With these operations, the high-frequency content of the planet - and possibly from the star, if it has a rich absorption lines content - is isolated from the speckles at the location of the planet. Note that the planet signal is still affected by telluric lines due to the atmospheric absorption.
Now that the planet is spectrally disentangled from the speckle contamination, molecular mapping \citep[][]{Snellen2015, Hoeijmakers2018} consists in cross-correlating, in the spectral dimension, the resulting signal $S_{\mathrm{res}}(\lambda,x,y)$ in each spaxel ($x$,$y$) from the IFS datacube with templates (see Fig. \ref{SNRgauss}) computed in regard to various companion properties (temperature, C/O ratio, metallicity, etc.). When cross-correlating a template with a spaxel corresponding to a companion, a correlation peak can be observed: its position corresponds to the radial velocity, and its width can be interpreted as being induced by the spin of the companion (\cite{cross_corr_snellen_2014}). 

The spectral template $t$ used for cross-correlation is user-defined, according to the type of planet atmosphere that is considered. It is expected that various templates will be tested on observation data, and they will come either from empirical libraries, or from theoretical models. A specific molecular spectrum can also be used to probe the presence of a molecule in the observed atmospheres \citep[][]{Hoeijmakers2018, ruffio_2019, Wang_2021, Cugno_2021}. Therefore, at a given radial velocity, the cross-correlation is just the scalar product of the residual signal, $S_{\mathrm{res}}$, and the Doppler shifted and normalized template, $t_{\mathrm{RV}}$, such that $\sum_{\lambda_i=\lambda_{\mathrm{min}}}^{\lambda_{\mathrm{max}}}t_{\mathrm{RV}}^{\,2}( \lambda_i) = 1$. This operation is described in Eq.\ref{eq:CCF}:
  
\begin{equation}
\begin{split}
\label{eq:CCF}
\mathcal{C}(v_{\mathrm{RV}},x,y) &=\sum_{\lambda_i=\lambda_{\mathrm{min}}}^{\lambda_{\mathrm{max}}} S_{\mathrm{res}}(\lambda_i,x,y) \cdot t(\lambda_i, v_{\mathrm{RV}}) \\
&= \langle S_{\mathrm{res}}(\lambda,x,y), t_{\mathrm{RV}} (\lambda) \rangle,
\end{split}
\end{equation}

where $v_{\mathrm{RV}}$ is the radial velocity, and $\langle A,B \rangle = \| A \| \, \| B \| \, \cos(\theta)$ denotes the scalar product of $A$ and $B$, and $\theta$ the angle between $A$ and $B$.

The expression of the correlation signal of interest is derived using Eq.\ref{residuals} with the approximate expression of the propagated noise $n(\lambda, x, y)$ and the definition of the scalar product:

\begin{equation}
\begin{split}
\label{scalarprod}
&\langle S_{\mathrm{res}} (\lambda,x,y), t_{\mathrm{RV}}(\lambda) \rangle \\
&\hspace{0.5em} \approx \langle\gamma_{\mathrm{atm}}(\lambda)\cdot[S_{\mathrm{planet}}(\lambda,x,y)]_{\mathrm{HF}}, t_{\mathrm{RV}}(\lambda)\rangle \\ 
&\hspace{1.5em}-\langle\gamma_{\mathrm{atm}}(\lambda)\cdot[\hat{S}_{\mathrm{star}}]_{\mathrm{HF}}(\lambda)\cdot\Bigg[\frac{S_{\mathrm{planet}}(\lambda,x,y)}{\hat{S}_{\mathrm{star}}(\lambda)}\Bigg]_{\mathrm{LF}}, t_{\mathrm{RV}}(\lambda)\rangle \\
&\hspace{1.5em}+ \langle n(\lambda,x,y), t_{\mathrm{RV}} \rangle \\[1.2em]
&\hspace{0.5em}\approx \Vert\gamma_{\mathrm{atm}}(\lambda)\cdot[S_{\mathrm{planet}}(\lambda,x,y)]_{\mathrm{HF}}\Vert \cdot \Vert t_{\mathrm{RV}} \Vert \cdot \cos(\theta_{\mathrm{planet,RV}})  \\
&\hspace{1.5em}-\Vert\gamma_{\mathrm{atm}}(\lambda)\cdot[\hat{S}_{\mathrm{star}}]_{\mathrm{HF}}(\lambda)\cdot\Bigg[\frac{S_{\mathrm{planet}}(\lambda,x,y)}{\hat{S}_{\mathrm{star}}(\lambda)}\Bigg]_{\mathrm{LF}}\Vert\cdot\Vert t_{\mathrm{RV}}\Vert \cdot \cos(\theta_{\mathrm{star,RV}}) \\
&\hspace{1.5em}+  \langle n(\lambda,x,y), t_{\mathrm{RV}}(\lambda) \rangle \\[1.2em]
&\hspace{0.5em}\approx \alpha(x,y) \cdot \cos(\theta_{\mathrm{planet,RV}}) - \beta(x,y) + \langle n(\lambda,x,y), t_{\mathrm{RV}}(\lambda) \rangle
\end{split}
\end{equation}

where $\displaystyle\theta_{\mathrm{planet,RV}}$ and $\displaystyle\theta_{\mathrm{star,RV}}$ are the angle between the high-pass filtered planet spectrum and the template, and between the high-pass filtered star spectrum and the template, respectively. Both depends on the input radial velocity. To simplify the notations, we have chosen to note the norm $\displaystyle\alpha = \|\gamma_{\mathrm{atm}}\cdot[S_{\mathrm{planet}}]_{\mathrm{HF}}\|$ and $\displaystyle\beta = \Vert\gamma_{\mathrm{atm}}\cdot[S_{\mathrm{star}}]_{\mathrm{HF}}\Bigg[\frac{S_{\mathrm{planet}}}{\hat{S}_{\mathrm{star}}}\Bigg]_{\mathrm{LF}}\Vert\cdot \cos(\theta_{\mathrm{star,RV}})$.

\begin{figure}
    \centering
    \includegraphics[width=9cm]{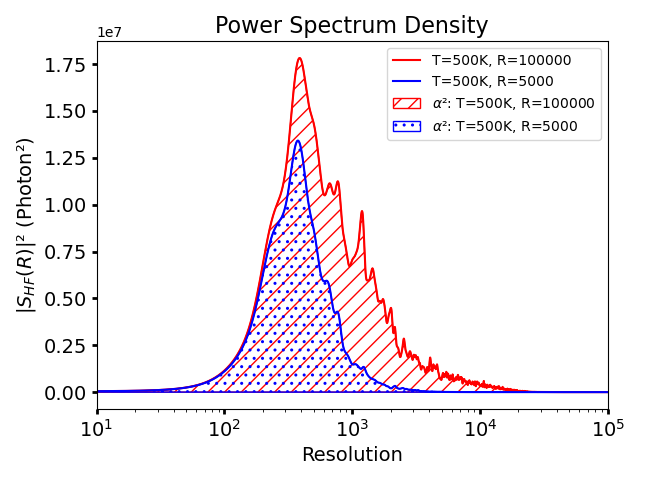}
    \caption{\label{DSP_vs_R} PSD of high pass filtered ($R_c=100$) planet spectrum (T=500K, log(g)=4, [M/H]=0) for two different resolutions on the total spectral range offered by BT-Settl spectra $(0.3 - \SI{12}{\micro m})$. The hatched areas correspond to the areas under the curves equal to the quantity $\alpha^2$.
    Note that the PSD has been smoothed only for better visualization purposes. The y-axis is relative to the total flux of the companions. One should only take into consideration the ratios between the curves explaining $\alpha$ variation from a given resolution to another one.}
\end{figure}

Hence, we have three terms in this measurement (Eq. \ref{scalarprod}). The first one - $\alpha$ - is the total planetary amount of information, including the transmission of telluric absorptions, after the high-pass filtering step. The cosine factor quantifies the similarity between the template and the companion spectrum so that $\displaystyle\alpha \cos(\theta_{\mathrm{planet,RV}})$ is the actual useful signal coming from the projection of the companion spectrum onto the considered template. The second one - $\beta$ - is the deterministic projection of the residual stellar spectrum lines onto the template. The last one is related to the noise propagated through the cross-correlation step. 

The detection is discussed according to the strength of the correlation peak value. This correlation peak is directly linked to both the $\alpha$ quantity, as it depends on the spectral content of the companion, and to the $\beta$ term that quantifies the rate of similarity between the star and the planet. This will be developed in subsection \ref{useful_information}. For the detection capability, this correlation peak has to be compared to the noise propagated onto the cross-correlation function (CCF, see Fig. \ref{SNRgauss}). This noise can be empirically estimated with the spatial variance of the correlation map or with the variance of the wings of the CCF (as long as the autocorrelation of the companion spectrum is taken into account to avoid biasing the noise estimate \citep[][]{Cugno_2021}). We will propose an analytical estimate of this propagated noise level (later detailed in subsection \ref{subsection:noise}), and we will check that this estimate is consistent with empirical estimation (in subsection \ref{validation}). 

\subsection{Molecular mapping useful information}\label{useful_information}

We focus here on the two first terms of Eq.\ref{scalarprod}, and detail the last one in the next subsection. Using the Parseval-Plancherel property of energy conservation in the Fourier domain, $\alpha$ can be expressed as the square root of the sum of the Power Spectrum Density (PSD) of the companion spectrum as in Eq.\ref{eq:alpha_freq}.

\begin{equation}
    \label{eq:alpha_freq}
    \begin{split}
    \alpha(x,y) &= \sqrt{\sum_{\lambda_i=\lambda_{\mathrm{min}}}^{\lambda_{\mathrm{max}}} \left( \gamma_{\mathrm{atm}}(\lambda_i)\cdot[S_{\mathrm{planet}}]_{\mathrm{HF}}(\lambda_i,x,y)\right)^2} \\
    &= \sqrt{\frac{1}{N_{\lambda}}\sum_{R_{i}=0}^{R_{\mathrm{max}}} \mathrm{FFT}\left\lbrace \gamma_{\mathrm{atm}}\cdot\left[S_{\mathrm{planet}}\right]_{\mathrm{HF}}(x,y)\right\rbrace^2(R_{i})} \\
    &= \sqrt{\frac{1}{N_{\lambda}}\sum_{R_{i}=0}^{R_{\mathrm{max}}} \textrm{PSD} \left\lbrace \gamma_{\mathrm{atm}}\cdot\left[S_{\mathrm{planet}}\right]_{\mathrm{HF}}(x,y)\right\rbrace(R_{i})} \, ,
    \end{split}
\end{equation}
where $N_{\lambda}$ is the number of spectral channels sampling the spectra, and the Fourier conjugate variable is associated with the spectral resolution $R$ because it is here homogeneous to $1/\lambda$.
\newline

Using the representation on the Fourier domain, Fig. \ref{DSP_vs_R} presents the PSD of the high-pass filtered spectrum for a companion with two different spectral resolutions, while keeping the same bandwidth (from \SI{1}{\micro m} to \SI{3}{\micro m}) and the same amount of photons. The area under the PSD curves is then equal to $\alpha^2$. In our study, this visualization allows us to better identify the spectral content location, and to directly compare it with the capabilities of an instrument with any given spectral resolution. Increasing the resolution expands the Fourier spectral range, and thus increases the signal of interest $\alpha$. Indeed, thin absorption lines in companion spectra induce a very high resolution content (even if the majority of the information seems - at first glance - to be found between a resolution of 100 and 10000). Having high resolution thus facilitates companion detection. In the same way, these representations in Fourier space allow us to visualize in Fig. \ref{DSP_vs_R_2} alpha dependencies as a function of planet temperature and spectral range.

\begin{figure}
    \centering
    \includegraphics[width=9cm]{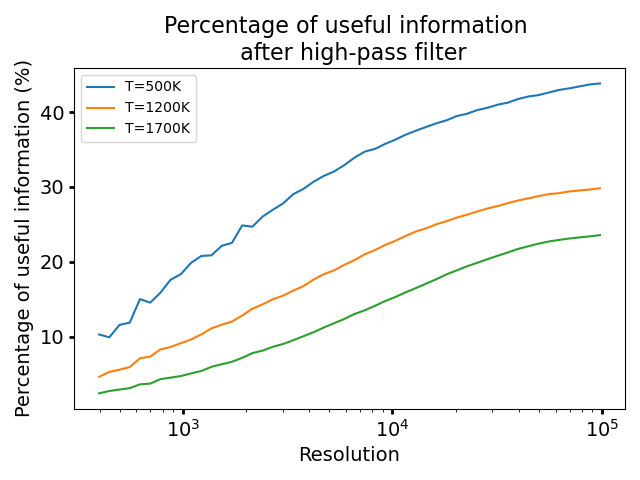}
    \caption{\label{fig:ratio_alpha} Fraction of the planetary spectrum remaining after high-pass filtering with a resolution cut $R_c=100$ as a function of the spectral resolution of the instrument (see text). Templates are chosen for different temperatures but with a constant gravitational surfaces (log(g)=4) and a constant metallicity ([M/H]=0). We notice that the colder the planet, the higher content of molecular absorption lines and the larger fraction of signal preserved at high resolution.}
\end{figure}

Therefore, this factor $\alpha$ quantifies the spectral richness of the companion spectrum and is homogeneous to photons. By analogy with ADI techniques where the signal of interest is the total number of photons coming from the planet, the signal of interest for molecular mapping lies in this factor. Note that - unlike ADI where the signal remains complete - because we are filtering out the low frequencies, we sacrifice about $60\%$ to $90\%$ of the planetary signal. We can define the fraction of the useful high frequency planetary spectrum over the total spectrum as described in Eq.\ref{equation:ratio_useful_information}. This quantity, as a function of the maximum resolution, is represented on Fig.\ref{fig:ratio_alpha}.

\begin{equation}
   \mathrm{F}_{\mathrm{planet}_{\mathrm{HF}}}(R_{\mathrm{max}}) = \frac{\sqrt{\sum_{R_{i}=0}^{R_{\mathrm{max}}} \textrm{PSD} \left\lbrace \left[S_{\mathrm{planet}}\right]_{\mathrm{HF}}\right\rbrace(R_{i})}}{\sqrt{\sum_{R_{i}=0}^{R_{\mathrm{max}}} \textrm{PSD} \left\lbrace S_{\mathrm{planet}}\right\rbrace(R_{i})}}
    \label{equation:ratio_useful_information}
\end{equation}

In other words, this ratio quantifies the cost of molecular mapping in terms of signal to be detected, with respect to ADI for instance, in order to get rid of the speckle noise limitation. 

As seen above, $\alpha$ can vary significantly depending on the planet temperatures (we considered a 500-1700K range in this study), as the high frequency spectral content is not the same.
$\alpha$ can be directly estimated with spectra models, resolution, and magnitude of the planet as inputs, and it does not require any end-to-end simulations. It depends on the companion spectrum, resolution, bandwidth, and exposure time. Note that the cosine factor is not estimated here, and we will assume it to be equal to 1 in the rest of the paper. But this could be a limitation of molecular mapping in real life since we have very few data to check the similarity between templates and real spectra. One should always keep in mind that differences likely exist between the template and the signal, and that the detection estimates that are derived when using identical information for the signal and the template are therefore computed in the most optimistic scenario.  

\begin{figure}
    \centering\includegraphics[width=9cm]{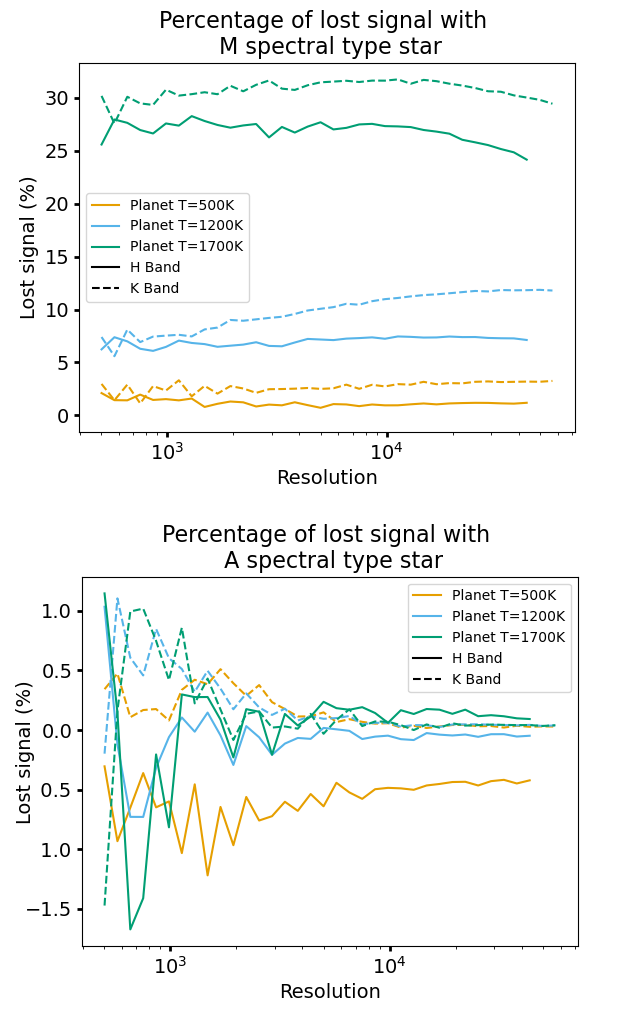}
    \caption{The percentage of information subtracted from the planetary signal due to the high-frequency content of the host star as a function of spectral resolution for several star and planet types. The stellar spectra that are used here come from the BT-NextGen database. The M-type star was chosen with a temperature of 3200K, and the A-type star with a temperature of 7200K. Both have a surface gravity log(g) = 4, and a metallicity [M/H]=0. This signal subtraction is obtained with non Doppler-shifted spectra to correspond to the worst case scenario where absorption lines common to both object are overlaid.}
    \label{fig:beta_soustraction}
\end{figure}

However, not all of this information may be usable for detection if the host star has lines in common with the planet. Indeed, subtracting the stellar spectrum will subtract those common lines as well. The term beta quantifies this effect, as it expresses the projection of the star's lines onto the template. Fig \ref{fig:beta_soustraction} shows the proportion of this self-subtraction for M- and A-host stars, and as a function of the planet type. M-stars and T planets have many CO lines, which explains the 30\% subtraction factor in K band with a 1700K planet. In contrast, A-stars have much less content in common with the companion spectra, and the subtractive term is almost zero. Note that beta can be negative if the cosine between the template and the stellar spectrum is negative, meaning that the spectra have a negative correlation. This explains the negative values in the figure. This negative correlation is very weak, however, as it has no strong physical origin. 

Also, various effects may induce some structure in the CCF as a function of radial velocity. The first effect is the intrinsic structure of the template autocorrelation. Radial velocity dependencies on the $\beta$ term is a secondary potential sources, even though it is minor in most of the cases. These facts are the primary motivation to assess the detection levels in the spatial dimension rather than in the radial velocity dimension. However, these effects can still affect the position of the peak as well as its shape. Therefore, one should take caution when deriving refined estimates of the companion orbital velocity or spin velocity.

\subsection{Noise}\label{subsection:noise}
The signal of interest, as quantified above, has now to be compared to the level of the noise term - $\sigma_{\mathrm{CCF}}^2$ - which is the variance of the projected noise onto the template $\langle n(\lambda,x,y), t_{\mathrm{RV}} \rangle$ in Eq. \ref{scalarprod}. It corresponds to the expected amount of noise in the correlation map (see Fig \ref{SNRgauss}). It is paramount to estimate this variance in order to later derive the analytical expression of the Signal-to-Noise Ratio (S/N), and to qualify a detection with a sufficiently high confidence level. 
Assuming that the PSF is centrosymmetric, the noise - $n(\lambda,x,y)$ - will only depend on $\lambda$, and on the separation $\rho$ as the star photon noise is the only source of variable noise in the FoV (Eq. \ref{eq:noise_breakdown}). Therefore, $\sigma_{\mathrm{CCF}}^2$ will depend on the separation $\rho$, and it can be decomposed with each noise contributor of Eq. \ref{eq:noise_breakdown}. We also make the assumption that, except for the speckle noise, the noise is uncorrelated at the pixel scale.

\begin{equation}
\begin{split}
    \sigma_{\mathrm{CCF}}^2(\rho) &= \mathrm{Var}[\langle n(\rho), t_{\mathrm{RV}} \rangle] \\
    &= \mathrm{Var}[\langle n_{\mathrm{halo}}(\rho), t_{\mathrm{RV}} \rangle + \langle n_{\mathrm{bkgd}}, t_{\mathrm{RV}}\rangle + \langle n_{\mathrm{RON}},t_{\mathrm{RV}} \rangle]  \\
    &= \sum_{\lambda_i=\lambda_{\mathrm{min}}}^{\lambda_{\mathrm{max}}} t_{\mathrm{RV}}^2(\lambda_i) \cdot \Big(\mathrm{Var}[n_{\mathrm{halo}}(\lambda_i, \rho)] + \mathrm{Var}[n_{\mathrm{bkgd}}(\lambda_i)] \\
    & + \mathrm{Var}[n_{\mathrm{RON}}]\Big)\\
    &= \sum_{\lambda_i=\lambda_{\mathrm{min}}}^{\lambda_{\mathrm{max}}} t_{\mathrm{RV}}^2 (\lambda_i) \cdot \sigma_{\mathrm{halo}}^2 (\lambda_i, \rho) + \sum_{\lambda_i=\lambda_{\mathrm{min}}}^{\lambda_{\mathrm{max}}} t_{\mathrm{RV}}^2 (\lambda_i) \cdot \sigma_{\mathrm{bkgd}} (\lambda_i)^2 \\
    &+ \sigma_{\mathrm{RON}}^2
    \label{noise_decomposition}
\end{split}
\end{equation}

To facilitate the reading the first two terms of Eq. \ref{noise_decomposition} will be noted as $\sigma_{\mathrm{halo, CCF}}^2(\rho)$ for the projected photon noise from the stellar halo at the separation $\rho$, and $\sigma_{\mathrm{bkgd, CCF}}^2$ for the projected photon noise from the sky and instrumental background, respectively. 

\subsubsection{Signal-to-Noise Ratio derivation}

The S/N at the location of the planet can be computed by comparing the signal of interest described earlier with the variance of the CCF values at the same separation as the planet (as shown in upper panels of Fig \ref{SNRgauss}). It can be derived as:
\begin{equation}
\begin{split}
    \textrm{S/N}(x_0,y_0) &= \frac{\alpha (x_0, y_0) \cos\left(\theta_{\mathrm{planet,RV}}\right) - \beta (x_0, y_0)}{\sigma_{\mathrm{CCF}}(\rho_{0})} 
    \label{SNR}
\end{split}
\end{equation}

where $(x_0, y_0)$ is the location of the planet and $\rho_0$ is the star-planet separation. Injecting the definition of the projected noise given in Eq.\ref{noise_decomposition} into Eq.\ref{SNR}, the S/N becomes:

\begin{equation}
\begin{split}
    \textrm{S/N}(x_0, y_0) = \frac{\alpha (x_0, y_0) \cos\left(\theta_{\mathrm{planet,RV}}\right) - \beta (x_0, y_0)}{\sqrt{\sigma_{\mathrm{halo, CCF}}^2(\rho_0) + \sigma_{\mathrm{bkgd, CCF}}^2 + \sigma_{\mathrm{RON}}^2}}  \\
    \label{SNR_sep}
\end{split}
\end{equation}

The signal of the planet being spatially distributed on several pixels, one can then integrate on a spatial box that is a few pixels wide (Eq. \ref{alpha_box}) to optimize the S/N as it is done with a matching template technique (\citet{ruffio_2019}).
\begin{equation}
    \alpha = \sum_{i=-w/2}^{w/2}\sum_{j=-w/2}^{w/2} \alpha (x_i, y_j),
    \label{alpha_box}
\end{equation}
where $w$ is the width of the box.

Assuming here that we cross-correlate planet spectra with a perfect template, the cosine would equal 1, and we can use Eq.\ref{SNR_sep} to derive the expression of the highest contrast that leads to a $5\sigma$ detection, as a function of separation, and for a given exposure time:

\begin{equation}
    \textrm{Contrast}(\rho) = \frac{5\sqrt{(\sigma_{\mathrm{halo, CCF}}(\rho)^2 + \sigma_{\mathrm{bkgd, CCF}}^2 + \sigma_{\mathrm{RON}}^2) \thinspace A_{\mathrm{FWHM}}}} {(\alpha_0-\beta_0)\sqrt{N_{\mathrm{exp}}}}
    \label{Contrast}
\end{equation}

where $\rho$ is the separation, $N_{\mathrm{exp}}$ is the number of exposures, $A_{\mathrm{FWHM}}$ is the area of the FWHM in pixels, $\alpha_0$ is the companion spectrum PSD integral over the FWHM area for a planet-to-star flux ratio equal to 1. The same applies to $\beta_0$ which is the projection of the template spectrum over the star spectrum for a planet-to-star flux ratio equal to 1. The typical FWHM size for HARMONI's apodized PSF is a bit more than 3 pixels wide. This core area covers 20\% to 30\% of the total flux from the planet depending on the AO correction.

\subsubsection{Discussion about the speckle noise} \label{speckle_noise}

The result obtained in Eq.\ref{residuals} is only valid if the estimation of the modulation functions (Eq. \ref{specmod_2}) is well done, and if the estimations of the star spectrum, the sky background spectrum, and the tellurics absorption are accurate. For the reasons mentioned above, we assumed that these modulations were on a finite frequency support, and that low-pass filtering could effectively separate the modulations from the high-frequency content of the companion. %Indeed, the speckle pattern is a function of wavelength and it can be described, at a first order, as a scaling effect through the FoV. In addition, and independently of this evolution, the Fresnel propagation of out-of-pupil aberrations cause the structure of the speckle field to slightly change with the wavelength. At a given spaxel position, this creates brightness variations across the wavelength dimension inducing a low-frequency modulation of the stellar continuum as mentioned above. 
However, if the frequency support of the modulations is not limited and/or a filter with a poor cut-off resolution is applied, residual modulations would result in another stellar residue (which would this time vary in the FoV), and add a new source of noise in Eq.\ref{noise_decomposition}. It is therefore important to check whether the assumption is valid by analyzing the PSD of the speckles modulation. This will also allow us to discuss the necessary cut-off resolution.

\begin{figure}[h]
    \centering
    \includegraphics[width=9cm]{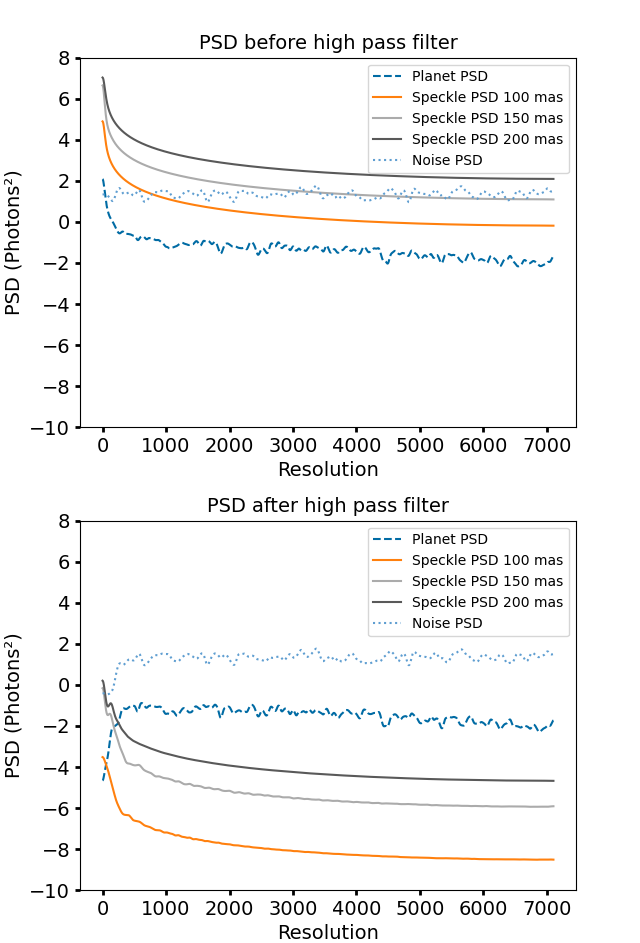}
    \caption{\label{PSD} The effect of high-pass filtering on speckles. Top panel: PSDs of the speckle modulation (taken at a several separations), photon and read-out noise and planet spectrum. Bottom panel: Same PSDs after the high-pass filtering with a resolution cut-off $R_c=100$.}
\end{figure}

Fig. \ref{PSD} shows the PSD of a speckle modulation taken at several angular separations, the white noise from photon and read-out noise, and the planet. For illustration purposes, we used the speckles derived from the system analysis of HARMONI, which are based on an end-to-end light propagation diffractive model of the instrument that reproduces their chromatic evolution. This reflects what one would expect to have with a good AO system, and partially calibrated non-common path aberrations (\cite{Carlotti2022a}, \cite{HCM_spec_2022}). 
We do not discuss here the relative levels between the curves, which strongly depend on the observation case, but we are here interested in the PSD shapes. 
The speckle modulation PSD has a large peak at low spectral resolution, and often dominates the others. This is why the speckle modulation has to be removed with a high-pass filter with a chosen cut-off frequency. In the same figure, the PSD of the speckle modulation still shows a high-frequency tail. This is because modulation of speckles is observed on a finite bandwidth window. The Fourier transform of this window is responsible for this PSD tail (i.e. the PSD of the speckle is convoluted by a sinc function corresponding to the Fourier transform of the rectangular function) dominating the intrinsic level of the speckle modulation. 

\begin{figure}[h]
    \centering
    \includegraphics[width=9cm]{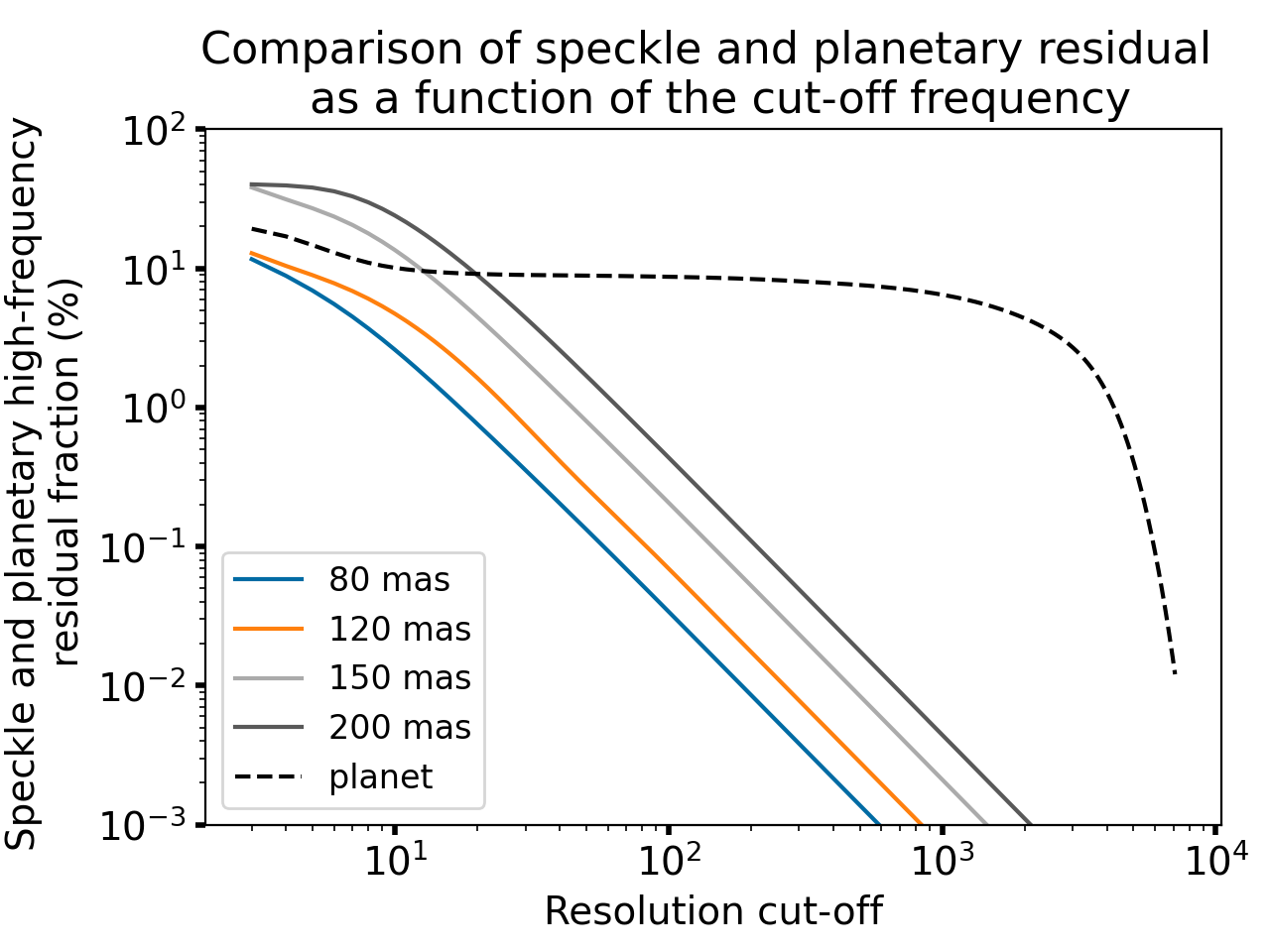}
    \caption{\label{Residuals} Residual fraction of the speckle modulation in respect to several angular separation (solid lines), and a planet spectrum (dashed line) as a function of the resolution cut-off. The planet spectrum is a BT-Settl template corresponding to T=1700K, log(g)=4, [M/H]=0 observed in K-band degraded to a spectral resolution of 7000.}
\end{figure}

To better visualize the frequency content, we propose to study the residual fraction of the speckle modulation and of a planet (T=1700K) as a function of the cut-off frequency of the high-pass filter (Fig. \ref{Residuals}). To do this, we use Eq.\ref{equation:ratio_useful_information}, this time fixing the resolution but varying the cutoff frequency of the high-pass filter. We notice that the fraction of residual speckle modulation increases with separation. This is well understood from the fact that the radial expansion rate of the speckles is proportional to the separation, and thus makes the modulations faster at larger separations. Would this parameter be critical, different cut-off frequencies could therefore be chosen depending on the separation. Note that on every curve, regardless of the separation, the main part of the speckle modulation is removed with a frequency cut corresponding to a spectral resolution of 100. This explains that on IFS like SPHERE and GPI (\cite{GPI_instru}), a high-pass spectral filter would be inefficient to disentangle the speckles modulation from the planet features. Having a spectral resolution greater than a few hundreds makes it feasible, however. It can also be seen that the planet spectra do indeed have a frequency content over a different frequency domain than the speckle modulations, as the planet PSD ratio decreases slowly as a function of the cut-off resolution.

Thus, frequency cut can be tuned up to optimize the S/N. A too high frequency cut would remove almost totally the speckles, but also the features of the planet, and thus degrade the S/N. On the contrary, a too low frequency cut would better preserve the planet, but also the speckle noise, degrading the S/N as well.  

\begin{figure}[h]
    \centering
    \includegraphics[width=9cm]{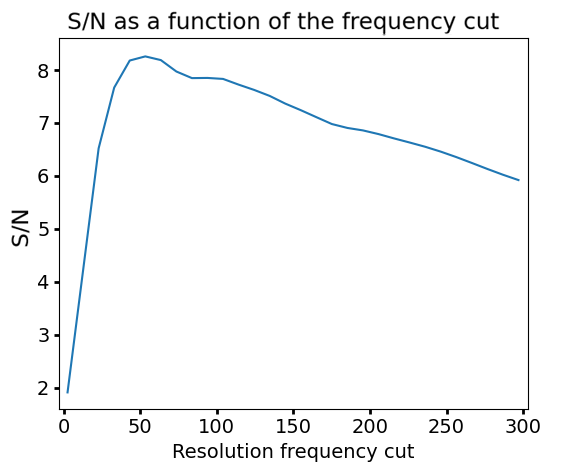}
    \caption{\label{speckle_filter}The plot shows the empirical S/N obtained with simulations for different cut-off frequencies obtained with the injection of a fake planet with the same HARMONI-like observation conditions as in Fig.\ref{PSD}.}
\end{figure}

\begin{figure}[t]
    \centering{}
    \includegraphics[width=9cm]{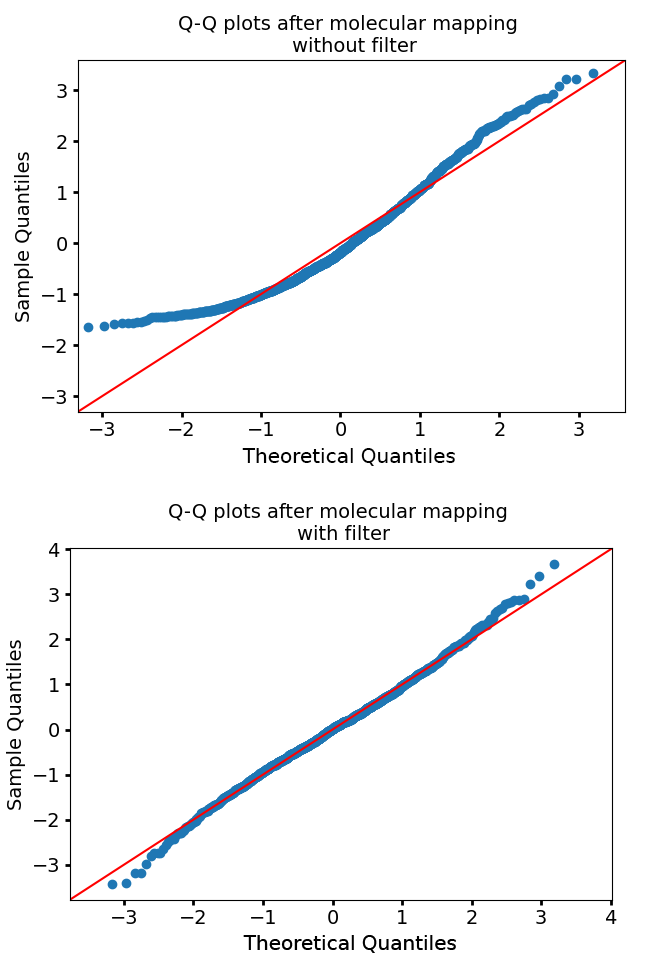}
    \caption{\label{qqplot} Study of noise statistics before and after high-pass filtering. Top panel: Quantile-quantile plot without speckle filtering. Bottom panel: Quantile-quantile plot with speckle high-pass filtering with a resolution cut-off of 100.}
\end{figure}

%This question about the ability to remove the impact of speckle noise as a function of the cut-off frequency can be addressed empirically through the test of fake companion detection in a simulated data cube.
Studying the ability to remove the impact of speckle noise as a function of the cut-off frequency can be done empirically through the test of fake companion detection in a simulated data cube.
Under the same astronomical conditions as in Fig. \ref{PSD}, Fig. \ref{speckle_filter} indicates an optimal cut-off frequency slightly above 50. 
As expected,  above a cut-off resolution of 100, speckles are almost entirely removed but the remaining planetary signal to be detected is slowly reduced. On the other hand, when the resolution cut-off is under 30, the influence of speckles is present, which degrades the S/N. For the following, we consider the case of a $R_c=100$ cut-off as a reasonable assumption. 
A second reason to choose a high enough cut-off resolution is the noise statistics. Indeed, every noise component is Gaussian except for the speckle noise, so, if the speckles are efficiently removed, the noise would be essentially Gaussian, and we could apply the ${5\sigma}$ threshold to the S/N to validate a detection with a $99.99994\%$ confidence level. The quantile-quantile plot in Fig. \ref{qqplot} is used to compare quantiles from the normal distribution with quantiles empirically estimated from the data. The closer the data are from the Gaussian distribution, the closer the blue dots are from the red line. It shows here that the spatial correlation matches well with a Gaussian distribution after the speckle suppression, and that the ${5\sigma}$ threshold can be applied.

The bottom panel in Fig. \ref{PSD} represents the PSD of the same components as in the top panel, but after the high-pass filtering. We see that the speckle noise is just under the photon and read-out noise. 
With this cut-off frequency, one millionth of the speckle PSD remains in the filtered spaxels. The term $\sigma_{\mathrm{speckles}}^2$ is therefore negligible compared to the other noises. We will therefore continue to neglect this noise in the following sections.\\ 

We will now check our S/N model with end-to-end simulations in various cases to conclude on the robustness of the validation.
 
\section{Test and validation of the S/N estimation model}
\label{section:validation}
%%%%%%%%%%%%%%%%%%%%%%%%%%%%%%%%%%%%%%%%%%%%%%%%%%%%%%%%%%%%%%%%%%%%%%%%%%%%%%%%%%%%%%

We propose to verify here our semi-analytical performance analysis in the specific case of the future ELT/HARMONI spectro-imager. We have a specific interest in this instrument as it is a major upcoming facility for the international community that will combine both a high-contrast imaging capability, and medium-to-high resolution spectroscopy. We revisit the high-contrast performance analysis for detections based on molecular mapping, and we compare the relative interest of various observing setups to provide useful information for the scientific preparation of future observations, and for the final adjustment of the instrument design. This study is also illustrative of other similar studies that can be applied to many other future high-contrast spectro-imagers. 

HARMONI is one of the 3 first-light instruments that are being developed for the ELT, ESO's 39-m segmented telescope. This visible and near-IR IFS is a general-use workhorse instrument that is designed to address a wide range of science goals \citep[][]{Thatte2020,Thatte2021,Thatte2022}. 

The high-contrast module (HCM) takes benefit from the combination of a good AO image quality on bright stars in the NIR, and spectral resolutions from 3500 to 17000. Using a dedicated fine correction of quasi-static instrumental aberrations (\cite{NDiaye2013,Hours2022}) and a coronagraph, it has been designed to characterize planets as close as 100 mas from their host star (goal 50 mas), and presenting a $10^{-6}$ flux ratio with it as described in \citet{Carlotti2022a}. It will allow interesting detections of the full - including continuum - spectrum of faint companions thanks to differential imaging, and we further discuss hereafter the additional capabilities of molecular mapping. 

\subsection{HARMONI high contrast mode properties}
The list of the possible spectral set-ups in the case of NIR high contrast observations are summarized in Table \ref{tableharmoni}. 

\begin{table}
      \caption[]{HARMONI spectral set-ups for high-contrast}
         \label{tableharmoni}
     $$ 
         \begin{array}{c c c c}%{0.5\linewidth}
            \hline
            \noalign{\smallskip}
            \textrm{Band} & \lambda_{\mathrm{min}} (\mu m) & \lambda_{\mathrm{max}}(\mu m) & \textrm{Resolution} \\
            \noalign{\smallskip}
            \hline
            \noalign{\smallskip}
            \mathrm{HK} & 1.450 & 2.450 & 3355 \\
            \hline
            \noalign{\smallskip}
            \mathrm{K} & 1.951 & 2.469 &  7104 \\ 
            \hline
            \noalign{\smallskip}
            \mathrm{H} & 1.435 & 1.815 & 7104 \\
            \hline
            \noalign{\smallskip}
            \mathrm{H_{high}} & 1.538 & 1.678 & 17385 \\
            \hline
            \noalign{\smallskip}
            \mathrm{K1_{high}} & 2.017 & 2.201 & 17385 \\
            \hline
            \noalign{\smallskip}
            \mathrm{K2_{high}} & 2.199 & 2.400 & 17385 \\
            \hline
         \end{array}
     $$ 
\end{table}

\begin{figure}
    \centering
    \includegraphics[width=9cm]{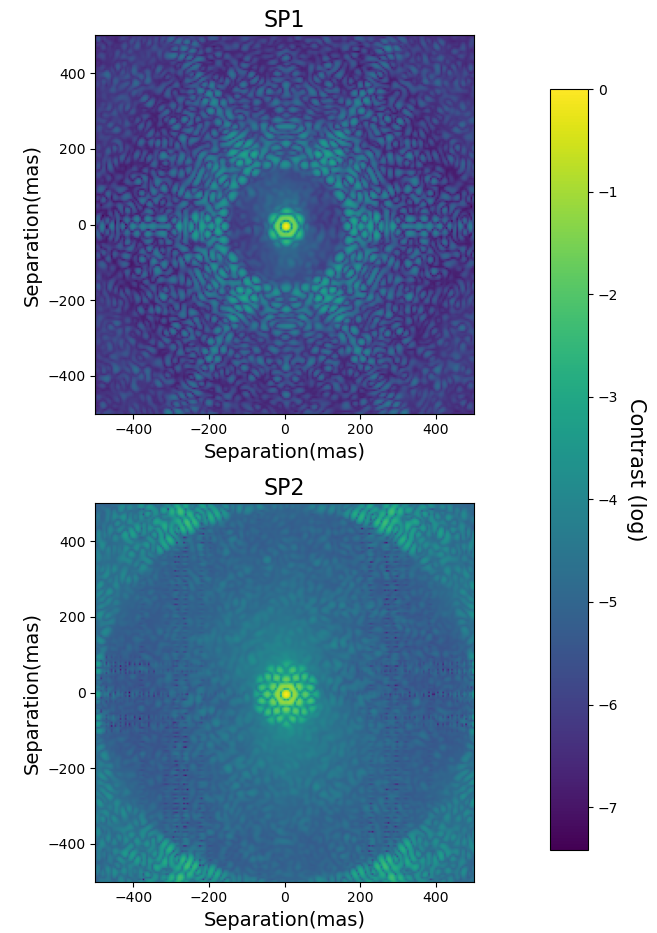}
    \caption{\label{PSF_apo}Monochromatic high contrast long exposure PSFs with the two HARMONI current apodizers. It has been simulated with the best seeing condition (Strehl ratio of 90\% in K-band).}
\end{figure}

Concerning image properties, we use the result of a realistic model of the apodized PSF of the telescope, as seen through the HCM. It considers the adaptive optics turbulence residuals (\citet{Neichel2016}) in the case of good / median seeing conditions for which high-contrast observations are considered, and that results in good image quality (Fig. \ref{PSF_apo}) with a Strehl ratio in K-band going from 85\% for median seeing to 95\% for the best conditions. The model also reproduces the non-common path aberrations by considering the specifications of the optics, and the way they are seen (or not) by the wavefront sensors, to reproduce the temporal and spectral evolution of the residual wavefront errors. The detailed end-to-end image formation simulation provides a long exposure time-averaged image (which is sufficient for the estimation of the main noise contributors in our approach) ; it also produces time-series that can feed an extensive end-to-end simulation (\cite{Houlle2021}). In this latter case, fake planets are injected to produce end-to-end contrast curves as a function of separation. We specify that, when comparing the obtained planet-to-star contrast in various spectral configurations, the contrast is defined here as the total flux ratio between the faintest detectable planet and the star over a fixed bandwidth from \SI{1.4}{\micro\metre} to \SI{2.5}{\micro\metre}. We argue that it is the best way to discuss the most favorable observation set-up for a given astronomical target.

In high-contrast observations, one of two coronagraph apodization patterns can be selected depending on the target brightness and separation range of interest (Table \ref{apo_table}). 

\begin{table}[h]
      \caption[]{Apodizers specifications.}
         \label{apo_table}
     $$ 
         \begin{array}{cccc}
            \hline
            \noalign{\smallskip}
            \textrm{Name} & \textrm{IWA  ($\lambda/D$)}  & \textrm{OWA ($\lambda/D$)} & \textrm{Transmission (\%)} \\
            \noalign{\smallskip}
            \hline
            \noalign{\smallskip}
            \textrm{SP1} & 5 & 11 & 45 \\
            \hline
            \noalign{\smallskip}
            \textrm{SP2} & 7 & 38 & 35 \\ 
            \hline
         \end{array}
     $$ 
\end{table}

Photon noise and read-out noise are added to background noise from the sky and the instrument estimated with the HARMONI simulator pipeline (HSIM)  (\citet{HSIM}).

We assume a detector read-out noise of 10 e-/pixel/read, typical of Teledyne's H4RG (\citet{TeledyneHRG}) for short detector integration times (DIT). Read-out noise can be attenuated if we use the "up-the-ramp" read mode of the detector. This mode allows taking $N_i$ intermediate images without resetting the pixels between the start and the end of the sensor exposure. Within the linear range, a linear regression provides an estimate of the total flux with a resulting read-out noise reduced by the square root of the non-destructive read-outs: $\sigma_{\mathrm{eff}} = \sigma_{\mathrm{RON}}/\sqrt{N_i}$. We take into account this read mode, and select $N_i$ according to the limiting DIT between the saturating DIT and the smearing DIT (i.e. the limit exposure time representing a displacement of the planet of a few tenths of $\frac{\lambda}{D}$). This smearing DIT appears in the case of pupil-stabilized observations. It depends on the separation of the planet, and on the position of the object in the sky. As a typical value, we are setting this DIT to 1 minute in our further considerations. This would roughly correspond to a minimum read-out noise of 4 e-/pixel/DIT. We also checked that this DIT value ensures that the background noise remains negligible with respect to read-out noise, even in K-band. 
From these assumptions, the method described in section \ref{molmapsection} allows to rapidly estimate the various noise contributors and resulting contrast curves for a wide diversity of targets and observation cases. This is implemented in the case of HARMONI within a dedicated package, named FastCurves, available on Github\footnote{https://github.com/ABidot/FastCurves}.

\subsection{Validation of FastCurves with respect to end-to-end simulations}\label{validation}

As a first and preliminary step, we want to validate our semi-analytical approach against end-to-end simulations in a specific case. In particular, this is aimed at checking the absolute values of the estimated useful signal of interest $\alpha$ and of the noise variance based on the halo brightness and detector noise, whereas the end-to-end simulation implements many realizations of noisy spectro-images, used as inputs to the molecular mapping signal extraction algorithm.
We use several test cases to try out the robustness of the method. We ran end-to-end simulations for different scenarii :

\begin{itemize}
  \item in the case of different types of planets and spectral range.
  \item in the case of an M-star with numerous absorption lines contaminating the planet spectrum.
  \item in the extreme case of a poor AO correction to test whether speckles are still not limiting.
\end{itemize}

\begin{figure*}
    \centering
    \includegraphics[width=18cm]{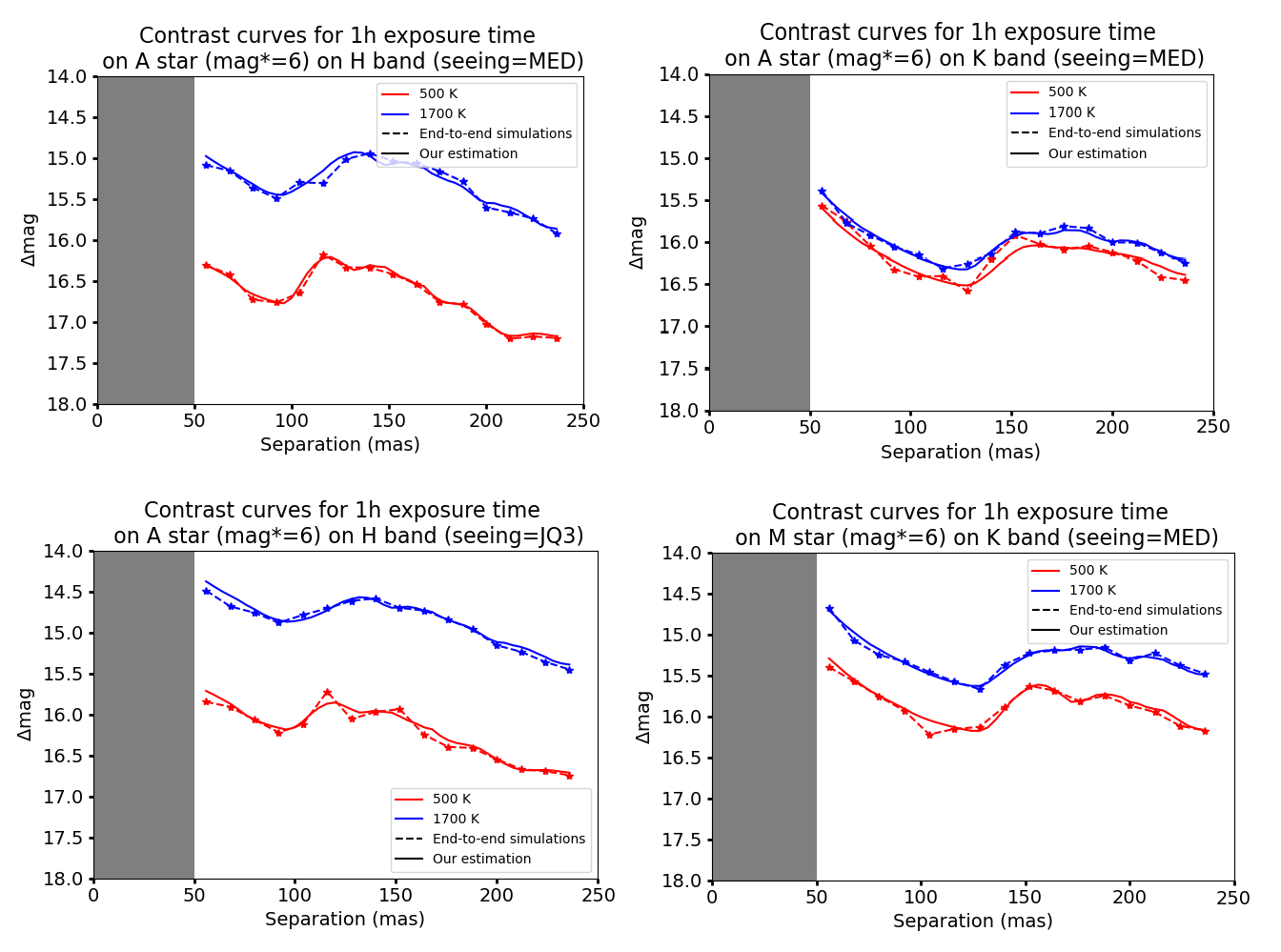}
    \caption{\label{end2end}Comparison of contrast curves obtained through our method with respect to end-to-end simulations. 
    The comparison has been made for 1 hour exposures with respect to several cases:   
    (top left panel) H-band HARMONI observation mode (Table.\ref{tableharmoni}) with an A-star (7200K, log(g)=4, [M/H]=0) magnitude of 6 with T-type planet (500K, log(g)=4, [M/H]=0) and L-type planet (1700K, log(g)=4, [M/H]=0) with median seeing condition (0.65'' for $\lambda=\SI{0.5}{\micro m}$, Strehl=94\%).
    (top right panel) same templates and conditions on K-band HARMONI observation mode. 
    (bottom right panel) same templates on H-band with JQ3 seeing condition (0.74'' for $\lambda=\SI{0.5}{\micro m}$, Strehl=73\%)
    (bottom left panel) same planetary templates but with an M-star (3200K, log(g)=4, [M/H]=0) on K-band with median seeing condition. 
    The solid lines are the prediction of FastCurves with these parameters, the dashed lines are the result of molecular mapping processing with time-series simulated PSF and fake planets injection.}
\end{figure*}

Fig. \ref{end2end} compares the contrast curves from our own end-to-end simulations and those from FastCurves. 
%The curves show a good agreement, with less than 0.3 magnitude difference in the worst case. 
Those test cases are an important validation of our general approach, and more specifically of the quantitative absolute estimates of the useful signal and noise. 
%It is relevant to easily produce relevant estimates of actual absolute SNR values for a variety of configurations, and not only some metrics to probe various trade-offs in a relative manner, as in the previous section. 
Finally, it also confirms the assumption that we made, i.e., that the residual speckle noise does become negligible after high pass-filtering it in the spectral dimension. 

%%%%Matthis discussion

We also compare our estimates to the results published by \citet{Houlle2021}. However, Houllé et al. used ATMO templates for fake planets injection and used BT-Settl for the cross correlation step, as a way to introduce some differences between the planet spectrum and the template spectrum. This leads to the combination of two effects affecting the estimation of detection performance: 
\begin{itemize}
\item a lower spectral content for ATMO resulting in a lower signal of interest $\alpha$,
\item a reduced similarity between templates, causing a cosine smaller than 1.
\end{itemize}
We took this into account to ensure the coherence of our work. We have computed the term $\alpha$ from the ATMO models, and we have estimated the mismatch with the cosine term which was computed as detailed in Eq.\ref{eq:cosine} :

\begin{equation}
    \cos(t_{\mathrm{ATMO}}, t_{\mathrm{BT-Settl}}) = \frac{\langle t_{\mathrm{ATMO}}, t_{\mathrm{BT-Settl}}\rangle}{\|t_{\mathrm{ATMO}}\| \cdot \|t_{\mathrm{BT-Settl}}\|}
    \label{eq:cosine}
\end{equation}

\begin{figure}[h]
    \centering
    \includegraphics[width=9cm]{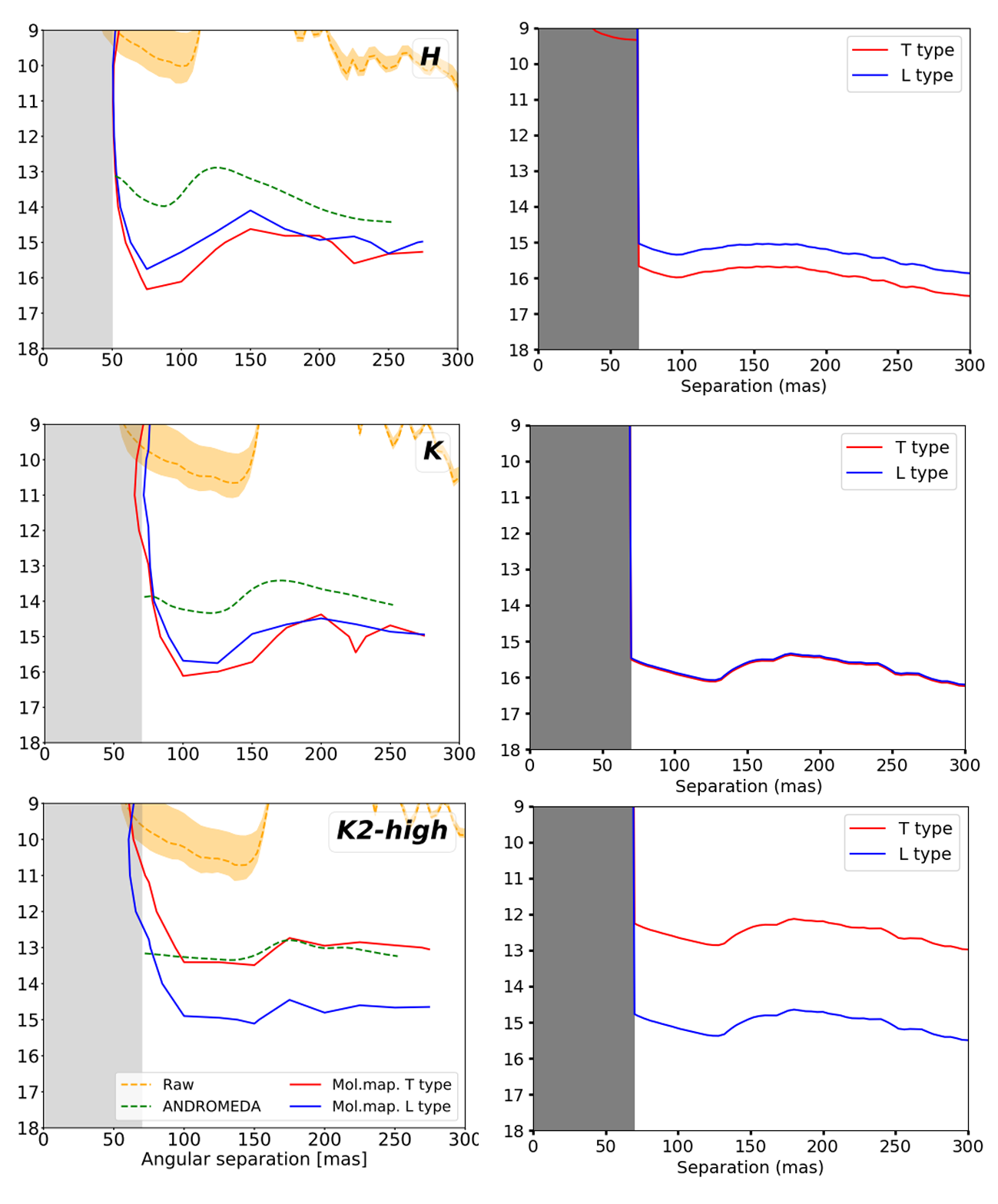}
    \caption{Comparison between our approach (right column) and results from Houllé et al. (left column). The comparison was made for the H, K and K2 high bands observation modes (see Table. \ref{tableharmoni}) with a star magnitude of 4.7, L and T spectral type companions, and a rotational broadening of 20 km/s. These curves are simulated with median seeing conditions with a Strehl ratio of about 80\% in K band. The total exposure time is set to 2 hours. Note that the green dashed curve on the left panel is the contrast curve obtained with the ADI ANDROMEDA algorithm (\cite{ANDROMEDA}), and the orange curve is the raw contrast with a standard deviation of 1$\sigma$.}
    \label{fig:comparaison_houllé_full}
\end{figure}

We find the same orders of magnitude as the authors did as illustrated in Fig. \ref{fig:comparaison_houllé_full}.

Note that relying only BT-Settl spectra, we would have estimated a 1-magnitude deeper contrast.

%\begin{figure}
%    \centering
%   \includegraphics[scale=0.22]{comparaison_houlle2.png}
%    \caption{\label{houlle_comp}Comparison of Houllé et al. results obtained with an end-to-end simulations (left panel), and FastCurves estimation (right panel). The comparison was made for the K-band observation mode (see Table. \ref{tableharmoni}) with a star magnitude of 4.7, L and T spectral type companions, and a rotational broadening of 20 km/s. These curves are simulated with median seeing conditions with a Strehl ratio of about 80\% in K band. The total exposure time is set to 2 hours. Note that the green dashed curve on the left panel is the contrast curve obtained with the ADI ANDROMEDA algorithm (\cite{ANDROMEDA}), and the orange curve is the raw contrast with a standard deviation of 1$\sigma$.}
%\end{figure}

\section{Trade-off for spectro-imaging}\label{trade_off}

We have seen that $\alpha$ (the amount of useful signal of the companion after continuum removal (Eq.\ref{eq:alpha_freq})) greatly depends on the resolution, the spectral range, and the type of planets. We want to explore this parameter space in regard to various trade-offs, and to evaluate the fluctuation of the S/N in the photon-limited case. Considering only the photon noise allows the study to take into account only relative values and  to not depend on both a particular star magnitude and integration time. We will discuss later the case of read-out noise in the HARMONI application case in section \ref{section:application_HARMONI}. 
 
\subsection{Quantified high-resolution signal of the exoplanet}

\begin{figure}[h]
    \centering
    \includegraphics[width=9cm]{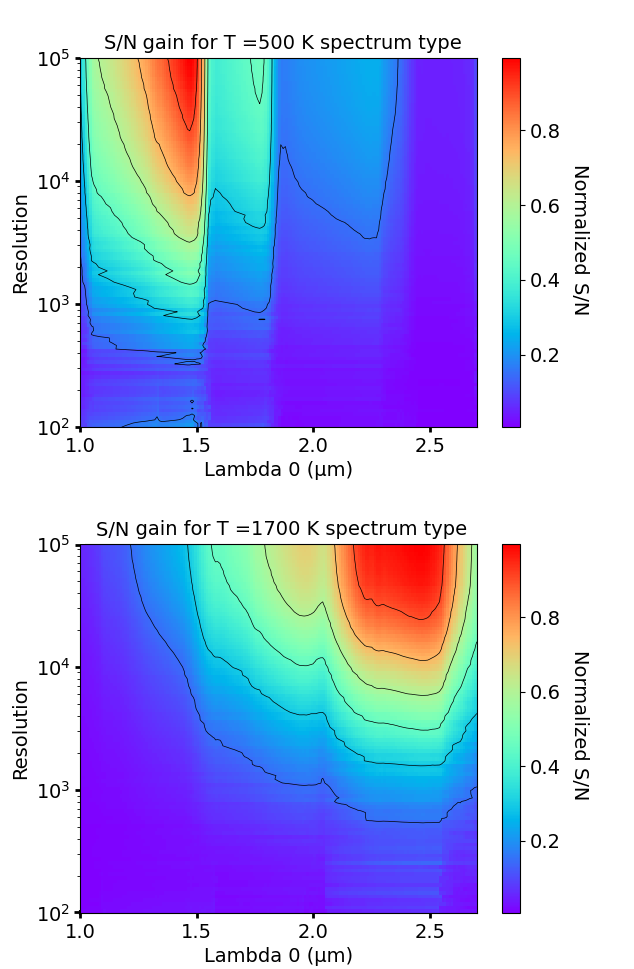}
    \caption{\label{SNR_no_tradeoff} S/N variation (defined in Eq.\ref{equation:SNR_no_trade_off}) as a function of the resolution and the central wavelength of the spectral range, with a constant \SI{0.5}{\micro\metre} large bandwidth. The $\alpha$ quantity has been estimated for a BT-Settl template corresponding to a temperature of 500K (top) and 1700K (bottom). The S/N has been normalized to the maximum (while the absolute values depend on many observing parameters, starting with the stellar magnitude and planetary contrast).}
\end{figure}

 Relying on BT-Settl spectra templates (\citet{Allard2012}), Fig. \ref{SNR_no_tradeoff} shows the variations of S/N for a T=1700K planet as a function of the resolution and the spectral range (corresponding to feasible ground-based observations), and assuming a constant \SI{0.5}{\micro\metre} large bandwidth, as described here in Eq.\ref{equation:SNR_no_trade_off}:
 
\begin{equation}
    \textrm{S/N}(\lambda_{0}, R_{\mathrm{max}})  = \sqrt{\frac{\frac{1}{N_\lambda}\sum_{R_{i}=0}^{R_{\mathrm{max}}}  \textrm{PSD}\left\lbrace \gamma_{\mathrm{atm}} \cdot S_{\mathrm{planet}, \lambda_{0}}\right\rbrace (R_{i})}{\sum_{\lambda_i=\lambda_{\mathrm{min}}}^{\lambda_{\mathrm{max}}} t_{\mathrm{RV}}^2 (\lambda_i) \cdot \sigma_{\mathrm{halo}}^2 (\lambda_i)}}
    \label{equation:SNR_no_trade_off}
\end{equation}

Where $\lambda_{\mathrm{min}}=\lambda_0 - \SI{0.25}{\micro m}$ and $\lambda_{\mathrm{max}}=\lambda_0 + \SI{0.25}{\micro m}$. \\

\begin{figure*}[h]
    \centering
    \includegraphics[width=18cm]{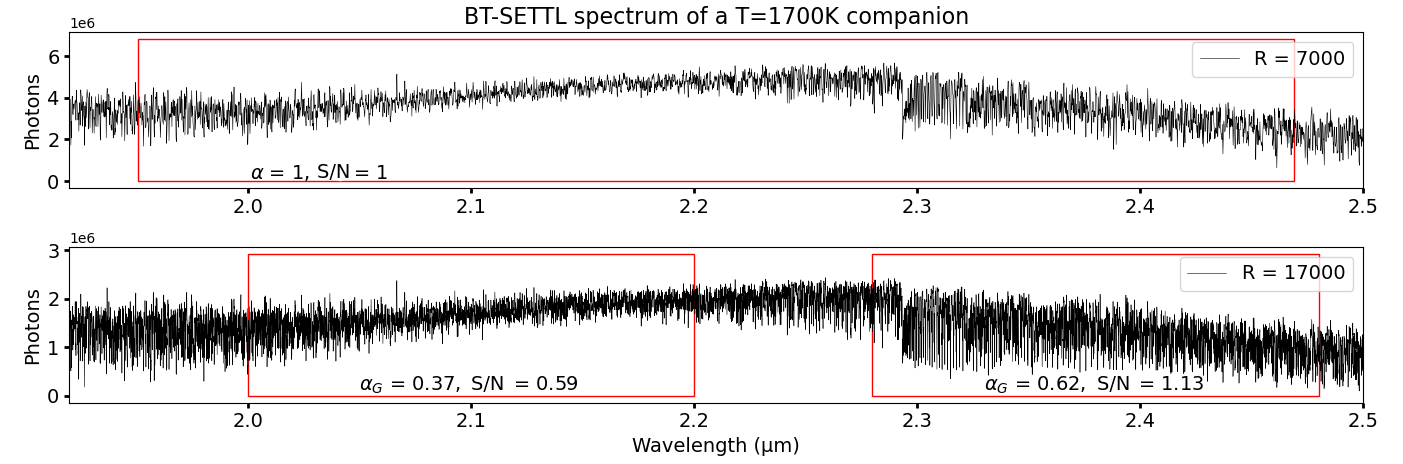}
    \caption{\label{fig:trade_off_spectrum}A BT-Settl spectrum of a T=1700K companion at two different spectral resolutions: R=7000 (top) and R=17000 (bottom). Each red rectangle delimits the observable spectral bands by keeping the same number of pixels for both resolutions. 
    In each case, the corresponding alpha quantities and the S/N, both normalized to the top panel, are indicated. 
    We note that some areas have more absorption lines, and thus benefit more than others from an increased resolution (for example, there is a forest of lines corresponding to the presence of CO around \SI{2.3}{\micro\metre}). 
    Both $\alpha$ and the photon noise decrease as the resolution increases and the bandwidth decreases; comparing the two variations may lead to a lower (left rectangle) or higher (right rectangle) resulting S/N. 
    } 
\end{figure*}

Note that here the noise factor, as mentioned above, only includes the propagated star photon noise contribution. We have chosen to take an A-star (T=7200K) to have a negligible $\beta$ term. The S/N maps have been normalized to only appreciate the variations of the S/N. 
The dependency over the resolution (along the y-axis) in Fig \ref{SNR_no_tradeoff} is directly related to Fig. \ref{fig:ratio_alpha}, that shows that the level of useful information increases with the spectral resolution. Quantitatively, in these examples, the gain in detectable signal can be as high as a factor 10 from low to high resolution. This can also be evidenced in Fig. \ref{fig:trade_off_spectrum} - which shows the same spectrum at two different spectral resolutions, and the corresponding high number of lines detected at higher resolution. 
In addition, we point out that this effect is quite inhomogeneous from a bandwidth to another, as there can be big S/N gain disparities for the same resolution. 

\subsection{Trade-off between bandwidth and resolution}

The limited pixel resources of a detector imposes a trade-off between resolution, bandwidth, and field of view (discussed later in subsection \ref{subsection:fov}). To respect the Nyquist criterion, a higher resolution mode requires more pixels to sample the spectra correctly and leads to a narrower spectral range. On the one hand, having a wider bandwidth provides a larger total number of photons and more companion absorption lines, which both increases $\alpha$. On the other hand, higher resolution allows for better separation of absorption peaks which become more numerous, deeper and more distinct, which also increases $\alpha$. 

This trade-off, under the constraint of a fixed number of pixels,  will obviously lead to different S/N variations than those presented in Fig. \ref{SNR_no_tradeoff} where the bandwidth was fixed and identical for any resolution. For example, in the previously introduced Fig. \ref{fig:trade_off_spectrum}, it can be seen that the value of $\alpha$ is higher at low resolution than at high resolution, but this factor is to be compared with the photon noise. Thus, a higher resolution coupled with a narrower bandwidth can result in either a better or a worse overall S/N, depending on the local wealth of spectral lines at high resolution, as illustrated in Fig. \ref{fig:trade_off_spectrum}. 

\begin{figure*}[h]
    \centering
    \includegraphics[width=18cm]{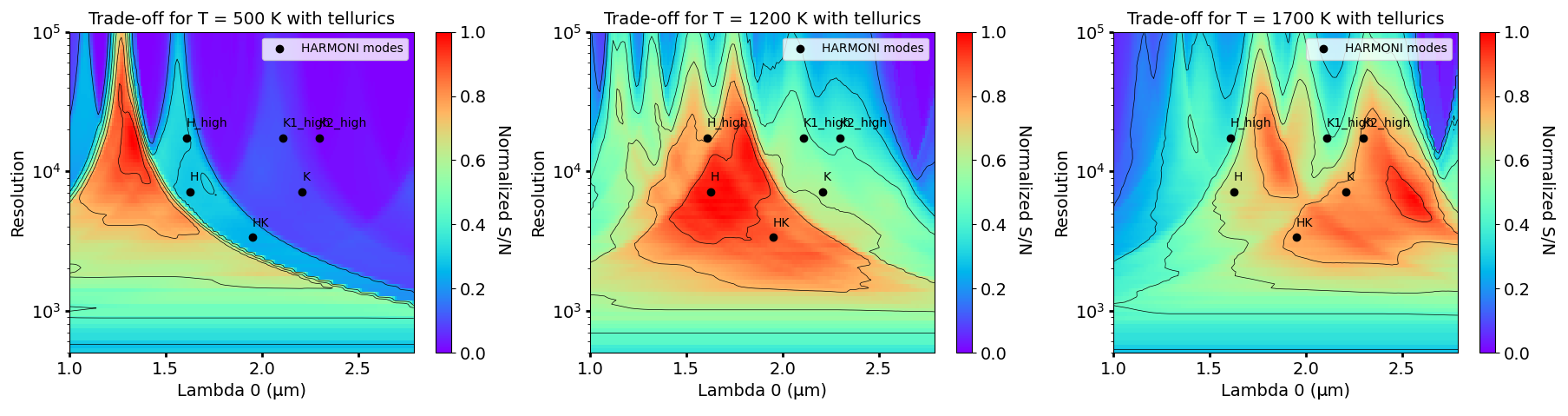}
    \caption{\label{SNRvsR}Evolution of the S/N (relative to its maximum) with the spectral resolution, for a fixed number of pixels (3300 pixels per spectrum here, corresponding to the case of HARMONI) and a correspondingly variable bandwidth. The largest spectral range is set to cover the H and K bands H (from \SI{1}{\micro\metre} to \SI{2.8}{\micro\metre}) for three different templates of planets signature. These maps include only the photon noise, not considering any wavelength-dependant instrumental transmission, which makes this result quite generic, informative of the planetary spectra properties and not of the instrument specificities. Three different planet temperatures are considered: 500 K (left), 1200 K (middle) and 1700 K (right).} 
\end{figure*}

Fig. \ref{SNRvsR} thus illustrates the variations of the S/N when assuming that the product of the resolution and of the bandwidth remains constant. Black dots represent the location of HARMONI IFU setups for future discussion (see Sec. \ref{subsection:curves}). 
We note that the higher the resolution, the more sensitive we are to the central wavelength. This is because the spectral range is limited, and therefore we must be well focused on the most interesting lines, and this is what induces the specific triangular shapes. Thus, choosing a too high spectral resolution could be risky. The red area corresponding to the optimal instrumental setup seems to point out a region where the resolution is around 8000 to 20000, and which moves with temperature from the H-band to the K-band.

Note that in the present trade-offs discussion, the detection boost is the only decision criterion. Obviously, the choice to go to higher resolution can also be motivated by the desire to have a better characterization of the physico-chemical properties of the planet in restricted bandwidths of interest.

\subsection{Spatial vs spectral information trade-off} \label{subsection:fov}

\begin{figure}
    \centering
    \includegraphics[width=9cm]{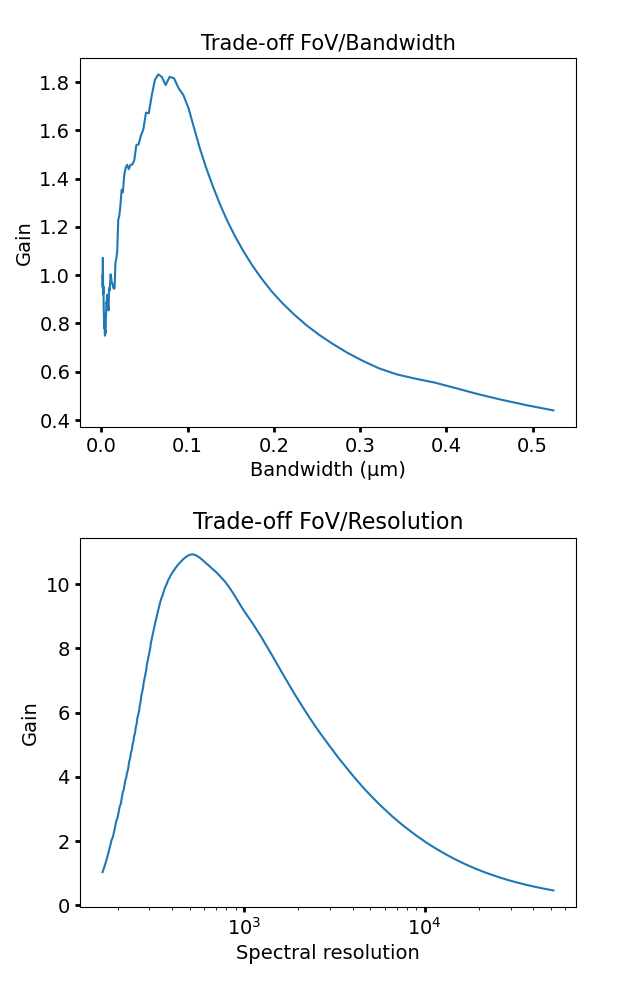}
    \caption{\label{fig:time_efficiency}Gain of time as a function of the bandwidth (top panel) or the spectral resolution (bottom panel) at the expense of mosaïcking the FoV for detecting a 500K planet (log(g)=4, [M/H]=0). The spectral range has been chosen to start at $\lambda_{\mathrm{min}} = \SI{1.2}{\micro m}$ for the FoV/Bandwidth trade-off and a fixed spectral range between $\SI{1.2}{\micro m}$ and $\SI{1.8}{\micro m}$ for the FoV/Resolution trade-off.}
\end{figure}

Another parameter involved in the trade-off discussion is the FoV. The larger the FoV, the fewer pixels will be available to either increase the resolution or increase the bandwidth. 
In the context of a blind search for a planet over a given FoV, one way to provide a quantitative metric for this trade-off is to compare the required total observation time needed to cover the whole FoV with a given S/N goal: either a larger FoV at once (with little spectral information Resolution$\times$Bandwidth, inducing a longer integration time per FoV), or mosaïcking over several smaller FoV, with richer spectral information. 
We show in Fig. \ref{fig:time_efficiency} the equivalent gain in observation time to reach the same detection level to cover a whole FoV by mosaïcking, either by increasing the resolution, or by increasing the bandwidth.

In these examples, above modest bandwidth ($>\SI{0.1}{\micro m}$), and resolution values (R>500), the gain of information increases less than linearly with the number of spectral pixels, and then a larger FoV is preferred over more spectral information. 
This general trend needs to be revisited in each specific case, however, and the proposed metrics makes it possible to account for any particular observation and planetary properties. This is also only valid for a blind search over the FoV. On the contrary, there may be external reasons to specifically limit the FoV of interest for molecular mapping. One obvious reason is when the location of the expected exoplanet is previously constrained. Another one is statistical, with a strong incentive to focus on small separations where exoplanets are expected to be more numerous, and where ADI approaches are more severely speckle-limited. 

\subsection{Additional interests in high resolution}
\subsubsection{Telluric absorption lines impact on detection}

\begin{figure}
    \centering{}
    \includegraphics[width=9cm]{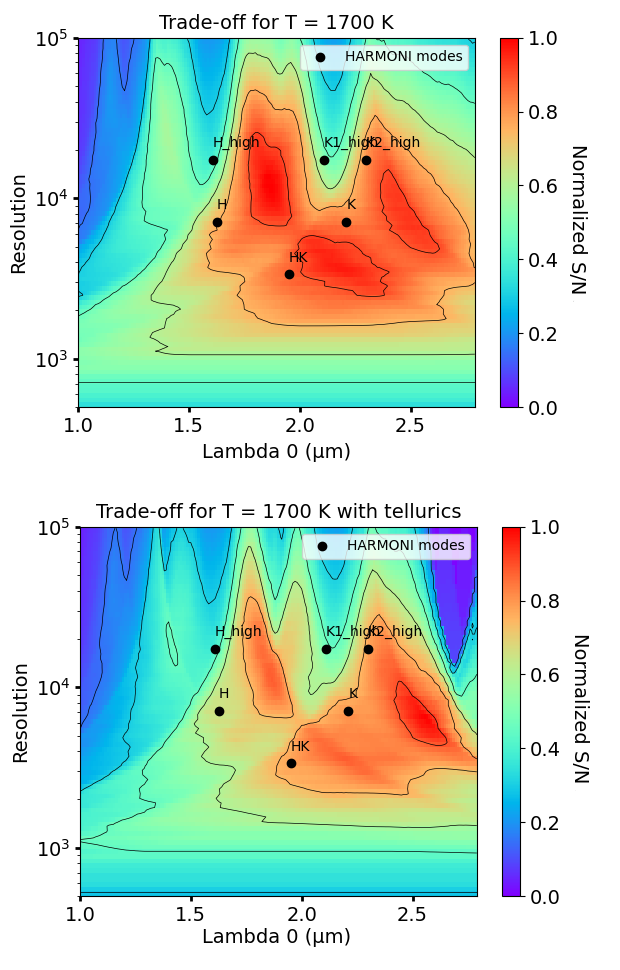}
    \caption{\label{tell_comparison} Comparison of S/N variations for a planet with temperature of 1700K with (bottom panel), and without (top panel) telluric absorption. The telluric absorption spectrum has been computed for an airmass of 1 using the ESO Skycalc module. The telluric absorptions do not impact the S/N distribution much, except that it makes it slightly more sensitive to the central wavelength, thus delimiting more clearly the spectral domains of interest. For instance, we see that near $\SI{2}{\micro m}$ the tellurics separate more the two zones of interest in H band and in K band by making the S/N vary more quickly.}
\end{figure}

\begin{figure}
    \centering
    \includegraphics[width=9cm]{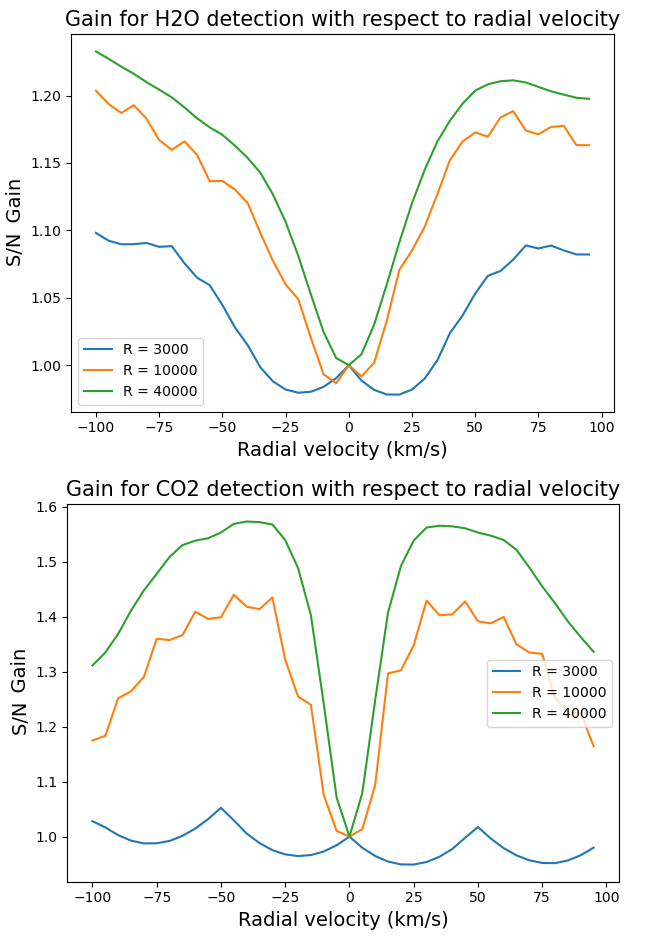}
    \caption{\label{rvshiftvstell}S/N gain as a function of radial velocity shift for H$_2$O and CO$_2$ molecules. The three curves represent this gain for different spectral resolutions (3000, 10000 and 40000).}
\end{figure}

According to Eq.\ref{residuals}, the telluric absorption still filters the spectrum of the planet after pre-processing. This can filter useful exoplanetary lines, resulting in a smaller value of $\alpha$. Fig. \ref{SNRvsR} already takes into account the transmission of tellurics in the estimation of $\alpha$ thanks to the ESO simulator Skycalc (\citet{skycalc}). Fig. \ref{tell_comparison} shows the impact of such tellurics on the S/N optimal zones. Despite a slight shift of these optimal zones, the impact of telluric absorption is moderate on young planets emitting their own light. This brings on average a degradation of the S/N by a factor of 0.9. This is because the correlation between the telluric absorption lines and those of the young companions is weak. 

However, the effect might be stronger when observing planets with an atmospheric composition more similar to the Earth. The correlation is much stronger, and our telluric absorption becomes troublesome by specifically removing the signal of interest from the exoplanet. 
Such a problem is partially mitigated if the exoplanet overall radial velocity shifts the spectral features, and if the spectral resolution is high enough to deblend the signal of interest from our atmosphere absorption. 
Relying on molecules opacities from the HITRAN database \citep[][]{HITRAN_database}, Fig. \ref{rvshiftvstell} illustrates this effect by estimating the gain in detection for spectral absorption lines caused by H$_2$O and CO$_2$ molecules as a function of the exoplanet radial velocity (with respect to Earth). 
As the resolution increases, the lines of interest appear in between the telluric absorption lines: a higher resolution allows a faster S/N gain (as a function of radial velocity) up to a high plateau value. The gain can even decrease as for some molecules, like CO$_2$, the absorption lines are almost evenly spaced in the spectral dimension. This is what causes the periodicity of the gain for a resolution of 3000 in Fig \ref{rvshiftvstell}.   

On the contrary, in the case of a young giant planet spectrum, there is no strong correlation with the exoplanet spectrum of interest, and we found no significant S/N gain with the radial velocity of the planet.

\subsubsection{Exoplanet reflected light vs stellar spectrum}
\begin{figure}
    \centering
    \includegraphics[width=9cm]{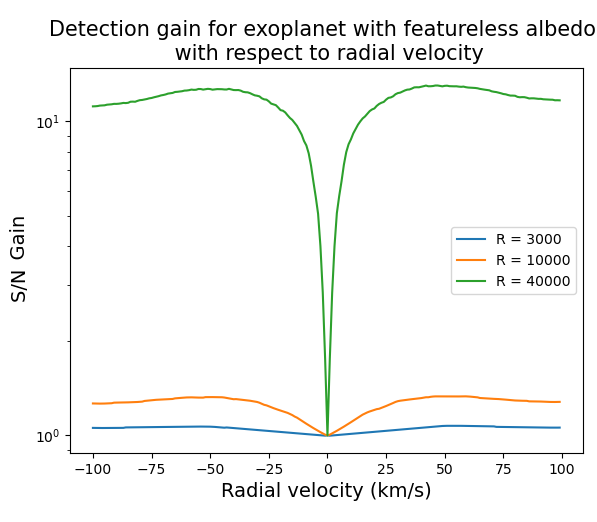}
    \caption{\label{fig:albedo_featureless}S/N gain as a function of a radial velocity shift for featureless exoplanetary albedo. The three curves represent this gain for different spectral resolutions (3000, 10000 and 40000).}
\end{figure}

If we consider the (worst) case of a featureless exoplanetary albedo, the exoplanet reflected light is essentially a fraction of the stellar spectrum shifted by the exoplanet orbital velocity (with respect to its host star). Despite the gray albedo, this Doppler shift will also enable molecular mapping to reveal the exoplanet within the stellar halo. Using a high-resolution spectrum of an M3 type star as an example (GJ388, as from the SPIRou database \citep[][]{Donati_2018}), we quantify the interest to go to higher resolution for this scientific case in Fig. \ref{fig:albedo_featureless}. This difficult case of high contrast reflected light exoplanets will definitely  push for high spectral resolution at the innermost separations.

\subsection{Rotational broadening impact on detection}

\begin{figure}
    \centering
    \includegraphics[width=9cm]{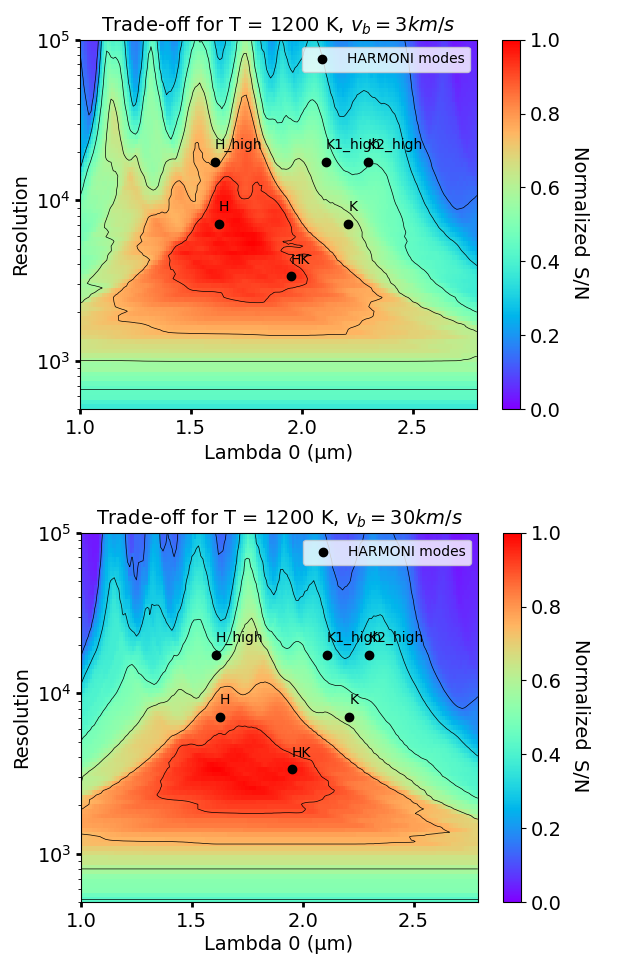}
    \caption{\label{SNRbroad} Evolution of the S/N with the resolution for the same exoplanet spectrum template (1200 K) but with two different rotational broadening corresponding to rotational speed of $3 km/s$ and $30 km/s$. The optimal area goes toward lower resolution as the rotational speed of the planet increases.}
\end{figure}

The exoplanet spin rotation induces a convolution of its intrinsic spectrum, broadening the specific lines. This attenuates the high frequency end of the spectrum PSD, and reduces the advantages of observing at higher resolution. This effect is observed in Fig. \ref{SNRbroad}, which compares colormaps representing the relative S/N of the same companion at a temperature of 1200K but with two different rotational velocities. As expected, the most interesting region shifts towards lower resolution values as the rotational velocity rises. 

%%%%%%%%%%%%%%%%%%%%%%%%%%%%%%%%%%%%%%%%%%%%%%%%%%%%%%%%%%%%%%%%%%%%%%%%%%%%%%%%%%%%%%
\section{Application to ELT-HARMONI high-contrast module} \label{section:application_HARMONI}

\subsection{FastCurves contrast predictions in various observing cases} \label{subsection:curves}
\begin{figure*}
    \centering
    \includegraphics[width=18cm]{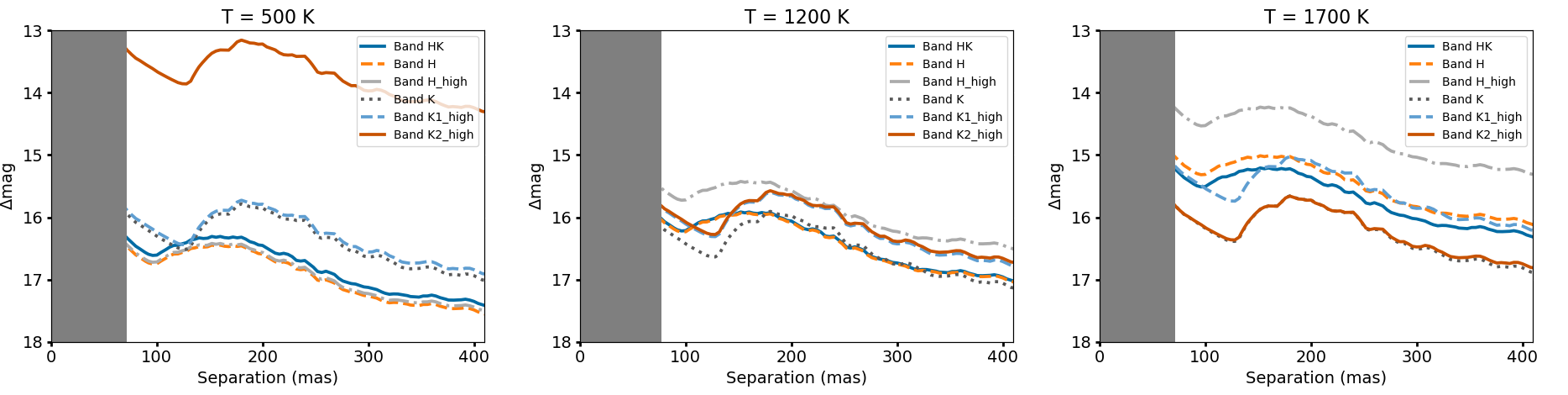}
    \caption{\label{fastcurves}Predicted contrast curves for HARMONI generated with the FastCurves package for three different planet temperatures. The exposure time is set to be 2h on a 6 magnitude A-star with a Strehl ratio of about 60\% in H band. }
\end{figure*}

We now explore a few cases of planet or companion types. The point here is not to be exhaustive but to cover a diversity of spectra cases, with very different effective temperatures with the BT-Settl model, from 500 to 1700K. We note that, for these cases, HARMONI modes mainly cover the most interesting part of the parameter space (Fig. \ref{SNRvsR}), and are complementary with respect to different planet temperatures and noise regimes. However, it suffers from not observing in J-band for the coldest companions. In each case, we present the obtained contrast curves for a 2 hours exposure and a 6th magnitude star, for various spectral setups (see Table \ref{tableharmoni}) with the apodizer SP1 (Table \ref{apo_table}). Results are displayed in Fig. \ref{fastcurves}.

These quickly generated contrast curves can be used as an exposure time calculator (ETC). They also help to determine the best observation mode that one should select according to the prior knowledge of the companions. 

 We find that the ultimate contrast achievable with the actual design seems to be a few $10^{-7}$ in two hours of effective integration (without including the system operation overheads). 
 As expected, a given contrast is more rapidly obtained on cool companions: warmer planets have wider and fewer lines, and thus a smaller signal to be detected in molecular mapping.  
 Also, T-type companions are more easily detected in the H band, unlike the L-type companions which are more easily detected in the K band.
 
In addition to the ability of the ELT to observe young planets as close as a few dozens of mas, it might enable the observation of planets in reflected light (\cite{Vaughan2022}).

\subsection{Discussion}
\subsubsection{Dominant noise regimes}\label{subsection:noise_HARMONI}
\begin{figure}
    \centering
    \includegraphics[width=9cm]{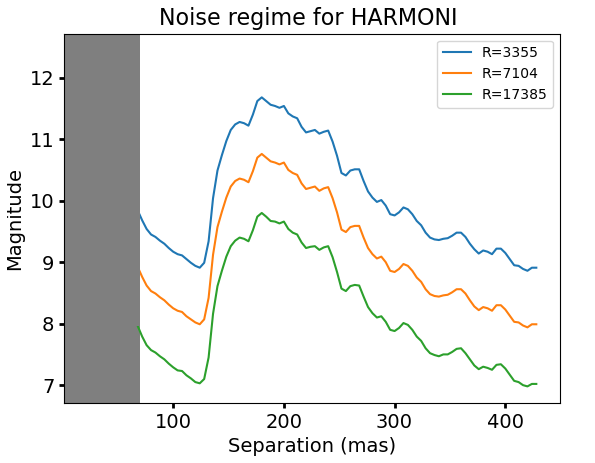}
    \caption{\label{noise_regime}Limit star magnitude above which photon noise becomes smaller than read noise as a function of separation and spectral resolution.}
\end{figure}

Since speckle noise is negligible in this approach, detection is either limited by photon noise, or by read-out noise. Fig. \ref{noise_regime} shows the dominant noise regimes as a function of the star magnitude, the separation, and the spectral resolution in the case of HARMONI. On this plot, we notice that for the high resolution mode (R$=$17385), the frontier between the photon noise limited case and the read-out noise limited case happens around a magnitude 8 in the short-separation, high-contrast region. 
This frontier is achieved thanks to the up-the-ramp detector read mode, which significantly reduces the impact of read-out noise. 
Unless the DIT can be significantly increased above 1 min, while remaining compatible with smearing and within the background-limited regime, these plots provide a good indication on the maximum spectral resolution to fully benefit from the collected photons and to avoid the read-out noise degradation of contrast. Quantitatively, these values will be slightly modified according to the image quality and the corresponding level of stellar halo. 

\begin{figure}
    \centering
    \includegraphics[width=10cm]{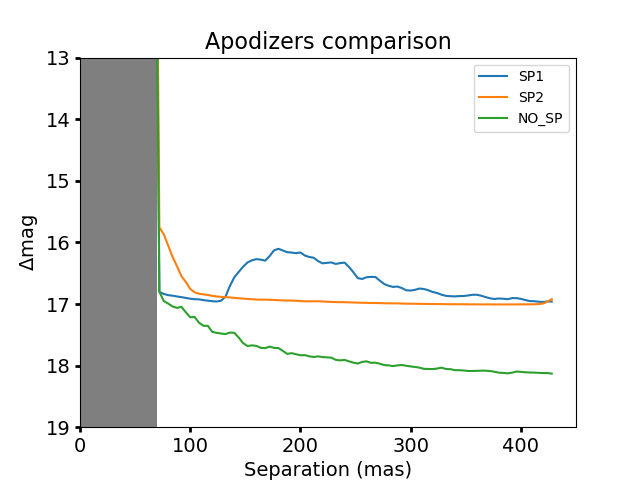}
    \caption{\label{contraste_apodizer}Contrast curves with and without apodizers. The comparison is made while considering the best seeing conditions (Strehl ratio of 95\% in K band).}
\end{figure}

\subsubsection{Revisit of the coronagraphic set-up} \label{corono_design}
Finally, we may revisit the definition and the selection of the optimal coronagraphic setup.
The high-contrast module includes two apodizers, designed for the more demanding assumption of ADI post-processing to recover the full companion spectrum. In such a case, it is interesting to minimize the intensity of residual diffracted light in the dark hole, not only to reduce the corresponding photon noise level, but also to limit any pinned speckles associated to the remaining diffraction pattern. This pushes the apodizers design towards higher contrast at the cost of transmission. 
This trade-off can be revisited in the case of molecular mapping where speckle noise is not anymore the main limitation\footnote{
End-to-end simulations confirm that, as previously (with apodizers), the PSD of the non-coronagraphic speckles has the same profile, and that a high pass filter removes them as well}.
In such a case, a new metric can be applied to design apodizers, comparing the transmission to photon noise, as detailed in 
Eq.\ref{apodizer_gain}.
\begin{equation}
    \frac{\textrm{S/N}_{\mathrm{apo}}}{\textrm{S/N}_{0}} = \sqrt{\frac{t_{\mathrm{apo}}}{t_0} \cdot \frac{\sigma_{0}^2}{\sigma_{\mathrm{apo}}^2}}  
    \label{apodizer_gain}
\end{equation}
where $\textrm{S/N}_{\mathrm{apo}}$ and $\textrm{S/N}_0$ are the S/N obtained respectively with and without apodizer. $t_{\mathrm{apo}}$ and $t_{0}$ are the throughput with and without apodizer. Finally, $\sigma_{\mathrm{apo}}$ and $\sigma_0$ are the brightness level of the adaptive optics halo for the two setups.

This new metric only takes photon noise into account.
In practice, FastCurves provides the comparison of various apodizer cases currently considered for HARMONI (see Fig. \ref{contraste_apodizer}). At short separation (100 mas), the non-coronagraphic case (no apodizer) actually outperforms the apodized setups with a 0.5 mag contrast gain. The gain becomes even larger at higher angular separation, with a gain of 1 magnitude. 

\section{Conclusions}\label{conclusion}

The proposed explicit expressions of the signal of interest and noise in the molecular mapping detection approach provide a very useful support for some trade-off studies in order to optimize future IFS instruments. 

We have confirmed the ability of molecular mapping to overcome the dominant limitation of speckle noise in ADI, especially at short separations. We also provide a quantitative estimate of the cost of this approach, which removes the exoplanet continuum and consequently reduces the signal to be detected. As expected, this cost strongly depends on the planet temperature. This approach also helps to discuss the relative interest of various spectral bandwidths as a function of spectral resolution and planet type. 

On top of the comparison of various setups relative performance, the application to the specific case of HARMONI, and the comparison to end-to-end simulations validate the estimate of absolute S/N values. The corresponding code FastCurves is accessible, and can be used as a direct ETC tool for HARMONI molecular mapping observations. 
These results provide some quantitative indications for the selection of the optimal set-up depending on the exoplanet type. It also predicts a high-contrast detection capability on bright targets of the order of a few $10^{-7}$ at 100 mas in 2h of exposure. 

FastCurves can also be easily adapted for similar studies in the case of other future instruments, by adjusting, in particular, the expected stellar profiles and the transmission. This approach is being used to propose an optimized design of the mid-resolution IFS considered for the VLT/SPHERE upgrade \citep[][]{Gratton_2022}.   

\begin{acknowledgements}
The authors thank the people in the HARMONI high-contrast science team, and in particular M. Bonnefoy, E. Choquet, and A. Vigan, for the exchanges they had with them.
This project is funded by the European Research Council (ERC) under the European Union's Horizon 2020 research and innovation programme (grant agreement n°866001 - EXACT).
\end{acknowledgements}

% WARNING
%-------------------------------------------------------------------
% Please note that we have included the references to the file aa.dem in
% order to compile it, but we ask you to:
%
% - use BibTeX with the regular commands:
%   \bibliographystyle{aa} % style aa.bst
%   \bibliography{Yourfile} % your references Yourfile.bib
%
% - join the .bib files when you upload your source files
%------------------------------------------------------------------

\bibliography{bibliography}
\bibliographystyle{aa} % style aa.bst
%\printbibliography

\begin{appendix}

\section{Frequency cut-off derivation}\label{appendixA}

The chosen Gaussian kernel for the convolution can be written as:
\begin{equation}
    G_{\sigma_c}(\lambda) = \frac{1}{\sqrt{2\pi\sigma_c^2}}\cdot\e^{-\frac{\lambda^2}{2\sigma_c^2}}
\end{equation}

We have the equality:

\begin{equation}
G_{\sigma_c}*S(\lambda) = \mathrm{FT}^{-1}\lbrace \mathrm{FT} \lbrace G_{\sigma_c} \rbrace \cdot \mathrm{FT}\lbrace S \rbrace \rbrace (\lambda) 
\end{equation}

We define the frequency cut-off such that:
\begin{equation}
\label{eq:freq_cut_equality}
    \mathrm{FT}\lbrace G_{\sigma_c} \rbrace (f_c)=\frac{1}{2}
\end{equation}

Using the usual Fourier Transform of a Gaussian we found that $\displaystyle \mathrm{FT}\lbrace G_{\sigma_c} \rbrace (f) = \e^{-2(f\pi\sigma_c)^2}$. Using Eq.\ref{eq:freq_cut_equality}, we derive the expression of the frequency cut-off as a function of $\sigma_c$:

\begin{equation}
    f_c = \frac{\sqrt{\ln{2}}}{\sqrt{2}\pi\sigma_c}
\end{equation}

The frequency cut-off $f_c$ is homogeneous to $1/\lambda$, and it can be translated into a spectral resolution cut-off as:

\begin{equation}
    R_c = \lambda_0\cdot f_c
\end{equation}
where $\lambda_0$ is the central wavelength of the spectrum.

\section{Stellar subtraction} \label{appendixB}

Here we seek to express both the residual planetary signal and the noise after molecular mapping preprocessing. Taking Eq.\ref{residuals} and developing it here, we obtain:

\begin{dmath}
    S_{\mathrm{res}}(\lambda,x,y) = S(\lambda,x,y) - \hat{M}_{\mathrm{speckle}}(\lambda,x,y)\cdot\hat{\gamma}_{\mathrm{atm}}(\lambda)\cdot \hat{S}_{\mathrm{star}}(\lambda)
    = \gamma_{\mathrm{atm}}(\lambda)\cdot\left(M_{\mathrm{speckle}} (\lambda,x,y).S_{\mathrm{star}} (\lambda) + S_{\mathrm{planet}} (\lambda,x,y)\right) + n(\lambda,x,y) - \hat{\gamma}_{\mathrm{atm}}(\lambda)\cdot\hat{S}_{star}(\lambda) \cdot \left(M_{\mathrm{speckle}}(\lambda,x,y) + \Bigg[\frac{S_{\mathrm{planet}}(\lambda,x,y)}{\hat{S}_{star}(\lambda)}\Bigg]_{\mathrm{LF}} + \Bigg[\frac{n(\lambda,x,y)}{\hat{\gamma}_{\mathrm{atm}}(\lambda).\hat{S}_{star}(\lambda)}\Bigg]_{\mathrm{LF}}\right)
    = \gamma_{\mathrm{atm}}(\lambda) \cdot S_{\mathrm{planet}}(\lambda,x,y) - \hat{\gamma}_{\mathrm{atm}}(\lambda)\cdot \hat{S}_{star}(\lambda) \cdot\Bigg[\frac{S_{\mathrm{planet}}(\lambda,x,y)}{\hat{S}_{star}(\lambda)}\Bigg]_{\mathrm{LF}} + n(\lambda,x,y)  - \hat{\gamma}_{\mathrm{atm}}(\lambda)\cdot \hat{S}_{star}(\lambda)\cdot\Bigg[\frac{n(\lambda,x,y)}{\hat{\gamma}_{\mathrm{atm}}(\lambda).\hat{S}_{star}(\lambda)}\Bigg]_{\mathrm{LF}}   
    \label{app_residuals}
\end{dmath}

Using the equality $S = [S]_{\mathrm{LF}} + [S]_{\mathrm{HF}}$, and assuming that the model of the tellurics transmission is perfect (i.e. $\hat{\gamma}_{\mathrm{atm}} = \gamma_{\mathrm{atm}}$), we find:

\begin{dmath}
S_{\mathrm{res}}(\lambda,x,y) = \gamma_{\mathrm{atm}}(\lambda) \cdot S_{\mathrm{planet}}(\lambda,x,y) - \gamma_{\mathrm{atm}}(\lambda)\cdot \left([\hat{S}_{star}]_{\mathrm{LF}}(\lambda)+[\hat{S}_{star}]_{\mathrm{HF}}(\lambda)\right) \cdot \Bigg[\frac{S_{\mathrm{planet}}(\lambda,x,y)}{\hat{S}_{star}(\lambda)}\Bigg]_{\mathrm{LF}} + n(\lambda,x,y) - \gamma_{\mathrm{atm}}(\lambda) \cdot \hat{S}_{star}(\lambda)\cdot\Bigg[\frac{n(\lambda,x,y)}{\gamma_{\mathrm{atm}}(\lambda).\hat{S}_{star}(\lambda)}\Bigg]_{\mathrm{LF}}
= \gamma_{\mathrm{atm}}(\lambda)\cdot\Bigg([S_{\mathrm{planet}}]_{\mathrm{HF}}(\lambda,x,y) - [\hat{S}_{\mathrm{star}}]_{\mathrm{HF}}(\lambda)\cdot\Bigg[\frac{S_{\mathrm{planet}}(\lambda,x,y)}{\hat{S}_{\mathrm{star}}(\lambda)}\Bigg]_{\mathrm{LF}}\Bigg) + n(\lambda,x,y) - [n(\lambda,x,y)]_{\mathrm{LF}} - \frac{[\gamma_{\mathrm{atm}}\cdot \hat{S}_{star}(\lambda)]_{\mathrm{HF}}}{[\gamma_{\mathrm{atm}}\cdot \hat{S}_{star}(\lambda)]_{\mathrm{LF}}} \cdot [n(\lambda,x,y)]_{\mathrm{LF}}
\end{dmath}

\begin{figure}[t]
    \centering
    \includegraphics[width=9cm]{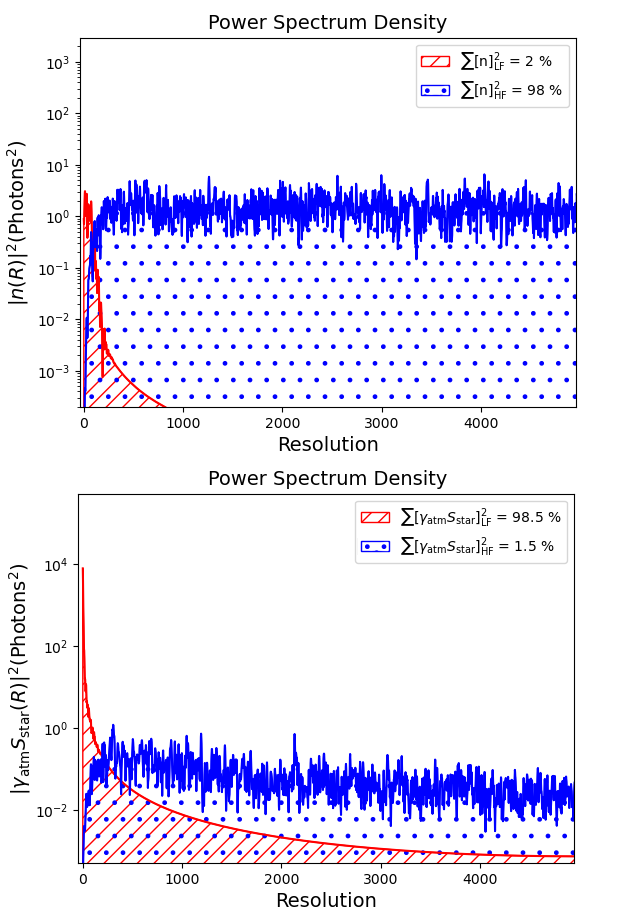}
    \caption{Relative weight of the low frequency and high frequency part of the noise and stellar signal. Top panel: PSDs of the high-pass filtered noise (in blue) and of the complementary low-pass filtered noise (in red). Bottom panel: Stellar spectrum PSD with the high-frequency part in blue and the low-frequency part in red. The star spectrum corresponds to a M-star and the tellurics to present a case where we have a strong high-frequency content.} 
    \label{fig:DSP_noise_star}
\end{figure}

While we could apply the full formula, it is convenient to highlight what are the dominant noise terms. We argue that, in usual cases, the dominant noise term is $n(\lambda,x,y)$. Indeed, the total variance of the noise is equal to $\displaystyle n^2(\lambda,x,y) + [n(\lambda,x,y)]^2_{\mathrm{LF}} + \frac{[\gamma_{\mathrm{atm}}\cdot \hat{S}_{star}(\lambda)]^2_{\mathrm{HF}}}{[\gamma_{\mathrm{atm}}\cdot \hat{S}_{star}(\lambda)]^2_{\mathrm{LF}}} \cdot [n(\lambda,x,y)]^2_{\mathrm{LF}}$. In practice, the filter cut-off frequency is kept small ($R_c \approx 100$) so that $\displaystyle [n]^2_{\mathrm{HF}} >> [n]^2_{\mathrm{LF}}$. It is however still large enough, considering the steep slope of the stellar DSP, so that $\displaystyle [\gamma_{\mathrm{atm}}\cdot \hat{S}_{star}]^2_{\mathrm{LF}} >> [\gamma_{\mathrm{atm}}\cdot \hat{S}_{star}]^2_{\mathrm{HF}}$. 
\newline
To illustrate these dominance relationships, Fig \ref{fig:DSP_noise_star} shows the PSD of these four components. Quantitatively, in the case studied in Fig \ref{fig:DSP_noise_star}, we find that the variance of the low-frequency noise, $\displaystyle [n]^2_{\mathrm{LF}}$, represents only 2\% of the high-frequency noise $\displaystyle [n]^2_{\mathrm{HF}}$. The fraction $\displaystyle \frac{[\gamma_{\mathrm{atm}}\cdot \hat{S}_{star}(\lambda)]^2_{\mathrm{HF}}}{[\gamma_{\mathrm{atm}}\cdot \hat{S}_{star}(\lambda)]^2_{\mathrm{LF}}}$ is also small, and in this example is equal to 0.015. Therefore, the noise term $n(\lambda,x,y)$ is the dominant one.

We thus find the expression of Eq.\ref{residuals}:

\begin{dmath}
S_{\mathrm{res}}(\lambda,x,y)\approx \gamma_{\mathrm{atm}}(\lambda)\cdot\Bigg([S_{\mathrm{planet}}]_{\mathrm{HF}}(\lambda,x,y) - [\hat{S}_{\mathrm{star}}]_{\mathrm{HF}}(\lambda)\cdot\Bigg[\frac{S_{\mathrm{planet}}(\lambda,x,y)}{\hat{S}_{\mathrm{star}}(\lambda)}\Bigg]_{\mathrm{LF}}\Bigg) + n(\lambda,x,y)
\end{dmath}

\section{Illustration of several signal dependencies}

\begin{figure}
    \centering
    \includegraphics[width=9cm]{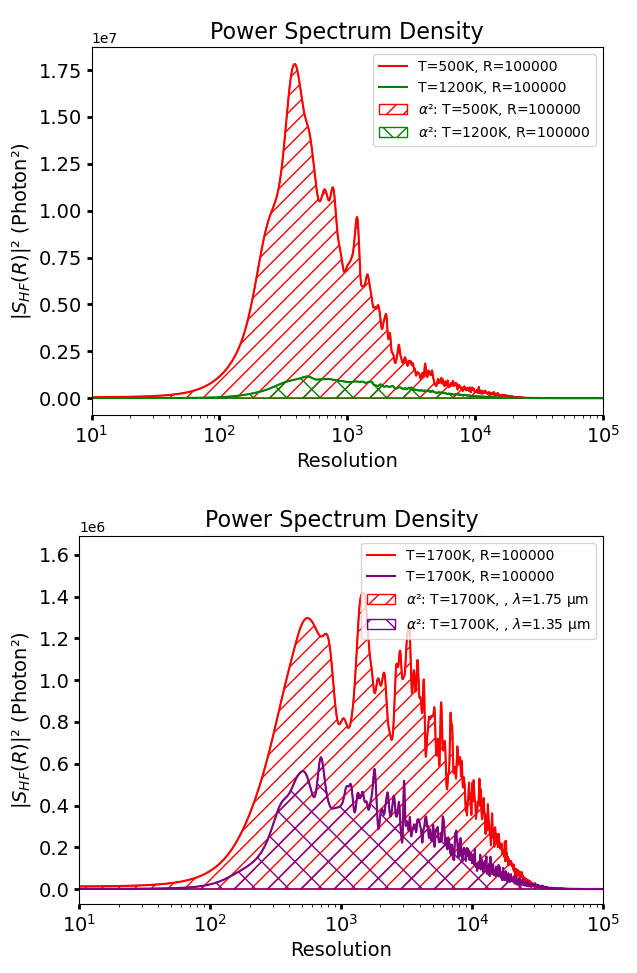}
    \caption{\label{DSP_vs_R_2} Visualization of the amount of information useful for molecular mapping detection. Top panel: PSD of two high pass filtered planet spectra (T=500K and T=1200K, log(g)=4, [M/H]=0) at the same resolutions and with the same amount of photons. Bottom panel: PSD of high pass filtered planet spectra (T=1700K, log(g)=4, [M/H]=0) on two different spectral ranges with a 300 nm bandwidth (center on $\lambda_0=\SI{1.35}{\micro m}$ in red and center on $\lambda_0=\SI{1.75}{\micro m}$ in purple).}
\end{figure}

As mentioned in the subsection \ref{useful_information}, the signal of interest depends not only on the spectral resolution of the instrument, but also on the intrinsic characteristics of the companion spectra. The top panel of Fig \ref{DSP_vs_R_2} quantifies this difference between a spectrum corresponding to an L-type planet and a T-type planet. Here, we find that for the same total photon count, the amount of signal is 6 times greater for a planet at 500K. 

Another significant dependency is the central wavelength of observation. The bottom panel of Fig \ref{DSP_vs_R_2} shows that we can obtain a difference of a factor of 1.6 for two central wavelengths spaced $\SI{400}{nm}$ apart. 

\end{appendix}
\end{document}